\documentclass[12pt]{article}

\usepackage{amsfonts}
\usepackage{amsmath}
\usepackage{amsthm}
\usepackage{cite}
\usepackage{epsfig}
\usepackage{latexsym}
\usepackage{paralist}
\usepackage{fancyhdr}
\usepackage{graphicx}
\numberwithin{equation}{section}
\usepackage[vcentermath]{youngtab}
\usepackage{young}
\usepackage{ytableau}
\usepackage{etex}
\usepackage{braket}
\usepackage{float}
\usepackage{autobreak}

\usepackage{pict2e}   
\usepackage{animate}  

\usepackage{multirow}
\usepackage{bigdelim}
\usepackage{fancybox}
\usepackage{cals}

\def\be{\begin{equation}}
\def\ee{\end{equation}}
\def\bea{\begin{eqnarray}}
\def\eea{\end{eqnarray}}

\renewcommand{\thefootnote}{\fnsymbol{footnote}}

\begin{document}

\hfuzz=100pt
\title{{\Large \bf{Exact results in 3d $\mathcal{N}=2$ $Spin(7)$ gauge theories \\with vector and spinor matters} }}
\date{}
\author{ Keita Nii$^a$\footnote{nii@itp.unibe.ch}
}
\date{\today}

\maketitle

\thispagestyle{fancy}
\cfoot{}
\renewcommand{\headrulewidth}{0.0pt}

\vspace*{-1cm}
\begin{center}
$^{a}${{\it Albert Einstein Center for Fundamental Physics }}
\\{{\it Institute for Theoretical Physics
}}
\\ {{\it University of Bern}}  
\\{{\it  Sidlerstrasse 5, CH-3012 Bern, Switzerland}}

\end{center}

\begin{abstract}
We study three-dimensional $\mathcal{N}=2$ $Spin(7)$ gauge theories with $N_S$ spinorial matters and with $N_f$ vectorial matters. The quantum Coulomb branch on the moduli space of vacua is one- or two-dimensional depending on the matter contents. For particular values of $(N_f,N_S)$, we find s-confinement phases and derive exact superpotentials. The 3d dynamics of $Spin(7)$ is connected to the 4d dynamics via KK-monopoles. Along the Higgs branch of the $Spin(7)$ theories, we obtain 3d $\mathcal{N}=2$ $G_2$ or $SU(4)$ theories and some of them lead to new s-confinement phases. As a check of our analysis we compute superconformal indices for these theories.
\end{abstract}

\renewcommand{\thefootnote}{\arabic{footnote}}
\setcounter{footnote}{0}

\newpage
\tableofcontents 
\clearpage

\section{Introduction}
Asymptotically-free gauge theories show various phases depending on the matter contents, the (global) structure of the gauge groups, spacetime dimensions, temperature and so on. It is usually difficult to exactly analyze the low-energy dynamics since it is strongly-coupled. In order to extract some analytic results, supersymmetry is a very useful tool. Non-renormalization theorems and holomorphy strongly constrain the SUSY dynamics and enable us to derive exact results \cite{Seiberg:1994bz,Seiberg:1994pq,Seiberg:1997vw}. For theories with four supercharges, supersymmetry can determine an exact form of the superpotential and we can find a quantum moduli space of vacua. In this paper, we are interested in low-energy dynamics of the supersymmetric $Spin(7)$ gauge theories.   

In four spacetime dimensions, $\mathcal{N}=1$ $Spin(N)$ gauge theories with vector matters and with various spinor matters were extensively studied (see \cite{Pouliot:1995zc,Cho:1997kr} \cite{Pouliot:1995sk,Pouliot:1996zh,Kawano:2007rz,Maru:1998hp,Strassler:1997fe,Berkooz:1997bb,Csaki:1997cu,Cho:1997sa,Kawano:1996bd, Csaki:1996zb}). For particular matter contents, the theories confine and the low-energy effective description has no gauge-interaction. For more general matter contents, we sometimes find the Seiberg dual descriptions which are the ``chiral'' theories and phenomenologically interesting.
In three spacetime dimensions, the corresponding $Spin(N)$ gauge theories are not well-studied. In \cite{Aharony:2013kma} (see also \cite{Aharony:2011ci}), the 3d $\mathcal{N}=2$ $Spin(N)$ guage theory with $N_f$ vector matters was investigated and its Seiberg duality was proposed by dimensionally reducing the 4d Seiberg duality. However the $Spin(N)$ gauge theories with spinor matters are not studied at all.

In this paper, we study the quantum aspects of the 3d $\mathcal{N}=2$ $Spin(7)$ gauge theories with spinorial and vectorial matters. Especially we will find new s-confinement phases for these theories and derive exact superpotentials which govern the confined phases. In order to verify the consistency of our analysis, we compute superconformal indices for these theories and for the dual (confined) descriptions. We will observe a complete agreement of the indices. As another check of our findings, we also test the various Higgs branch. Along the Higgs branch we find the s-confinement description of the 3d $\mathcal{N}=2$ $G_2$ or $SU(4)$ gauge theories with various matters. For the $G_2$ Higgs branch, we will reproduce the same superpotential discussed in \cite{Nii:2017npz}. Along the $SU(4)$ Higgs branch, we reproduce the known s-confinement phases and also find new s-confinement phases for the 3d $\mathcal{N}=2$ $SU(4)$ gauge theories with anti-symmetric matters. We also discuss the connection to the 4d $\mathcal{N}=1$ $Spin(7)$ gauge theories by incorporating the KK-monopoles.

This paper is organized as follows. 
In Section 2, we will briefly review the 4d $\mathcal{N}=1$ $Spin(7)$ gauge theories with spinorial matters. 
In Section 3, the Coulomb branch of the $Spin(7)$ vector multiplet is (semi-)classically studied.  
In Section 4, a 3d $\mathcal{N}=2$ $Spin(7)$ gauge theory with matters in a spinorial representation is investigated. We also compute the superconformal indices. 
In Section 5, we study the $Spin(7)$ theory with spinor and vector matters with special attention to the s-confinement phases. 
In Section 6, we summarize our results and comment on possible future directions.

\section{Review of 4d $\mathcal{N}=1$ $Spin(7)$ gauge theories}
In this section, we will briefly review the dynamics of the 4d $\mathcal{N}=1$ $Spin(7)$ gauge theories with spinorial matters. Table \ref{4dspin7} shows the matter contents and their quantum numbers. Due to the chiral anomalies in 4d, $U(1)$ and $U(1)_R$ global symmetries are anomalous and then we have to combine them into a new $U(1)_R$ symmetry
\begin{align}
U(1)_R^{new}=U(1)_R^{old} -\left(R_S-1 +\frac{5}{N_S}   \right)U(1).
\end{align}
In this paper, we are interested in the 3d theories and these $U(1)$ symmetries are not anomalous. Hence we will use spurious charge assignment also in 4d.

\begin{table}[H]\caption{Quantum numbers for 4d $\mathcal{N}=1$ $Spin(7)$ theories} 
\begin{center}
  \begin{tabular}{|c||c||c|c|c|c|c| } \hline
  &$Spin(7)$&$SU(N_S)$&$U(1)$&$U(1)_R$ &$U(1)_R^{new}$ \\ \hline 
$S$ & $\mathbf{2^{N}}=\mathbf{8}$&${\tiny \yng(1)}$&1& $R_S$&$1-\frac{5}{N_S}$ \\
$\lambda$ &$\mathbf{Adj.}$&1&0&$1$  &1\\ \hline
$\eta=\Lambda_{N_f,N_S}^b$&1&1&$2N_S$&$2N_S(R_S-1) +10$ &0 \\ \hline 
$M_{SS}:=SS$&1&$\tiny \yng(2)$&2&$2R_S$ &$2-\frac{10}{N_S}$\\
$B_S:=S^4$&1&${\tiny \yng(1,1,1,1)}$&$4$&$4R_S$  & $4-\frac{20}{N_S}$ \\[10pt] \hline
  \end{tabular}
  \end{center}\label{4dspin7}
\end{table}

\noindent In Table \ref{4dspin7}, $\eta$ is a dynamical scale of the $Spin(7)$ gauge group and $b$ is a coefficient of the one-loop beta function, which is given by 
\begin{align}
\beta &=-\frac{g^3}{16  \pi^2} b,~~~b=15-N_S.
\end{align}

Quantum dynamics depends on the number of spinor multiplets. We simply enumerate the results and give some comments. For $N_S \le 3$, we only need the gauge invariant $M_{SS}$ in order to describe the Higgs branch. The superpotential to govern the low-energy dynamics is
\begin{align}
W_{N_S \le 3} &=\left(\frac{\eta}{\det M_{SS}}  \right)^{\frac{1}{5-N_S}} ~~~~(N_S \le 3).
\end{align}
From the superpotential, there is no stable SUSY vacuum. At generic points of the moduli space, the gauge group is maximally broken to $SU(2)$. The gaugino condensation of the remaining $SU(2)$ generates this superpotential.

For $N_S=4$, we need also the baryonic operator $B_S$. At generic points of the moduli space, the gauge group is now completely broken and thus we can reliably use the instanton calculation. One-instanton configurations generate 
\begin{align}
W_{N_S=4} &=\frac{\eta}{\det M_{SS} -B_S^2}.
\end{align}
For $N_S=5$, the Higgs branch coordinates $M_{SS}$ and $B_S$ need one constraint between them. The classical constraint is modified quantum-mechanically and realized by using the Lagrange multiplier $X$ as 
\begin{align}
W_{N_S=5} &= X\left(  \det M_{SS} -M_{ij}B^i B^j -\eta \right).
\end{align}
For $N_S=6$, the quantum moduli space is the same as the classical one. The classical constraints between the Higgs branch coordinates are depicted as
\begin{align}
W_{N_S=6} &=\frac{1}{\eta} \left( \det M -M_{ik} M_{jl} B^{ij}B^{kl} -\mathrm{Pf}\, B  \right). \label{4dF6}
\end{align}
For the 4d $\mathcal{N}=1$ $Spin(7)$ gauge theory with spinors and vectors, we will not review it here and see \cite{Cho:1997kr, Csaki:1996zb}.

\section{Coulomb branch and Monopole operators}
In this section, we will define the (semi-)classical Coulomb branch coordinates which correspond to the monopoles with a magnetic charge $g_i=\vec{\alpha}^*_i \cdot \vec{H}$, where $\vec{\alpha}_i$ is a simple root and $\vec{\alpha}^*_i $ denotes a dual root $\frac{2 \vec{\alpha}}{\vec{\alpha}^2}$. $\vec{H}$ is a Cartan subalgebra. The Coulomb branch operators parametrize the flat directions of the scalar fields from the vector superfields. The adjoint scalar field in a vector superfield is defined as
 \begin{align}
\phi := \left( \sum_{i=1}^r \phi_i \vec{\alpha}^*_i \right)  \cdot  \vec{H},
\end{align}
where we used the gauge transformation and diagonalized the adjoint scalar into the Cartan part. In this notation, the weyl chamber is given by
\begin{align}
\sum_{j=1}^r \,A_{ij}  \phi_j  \ge 0 ~~~(\text{for each } i)
\end{align}
where $A_{ij}:=\vec{\alpha}_i \cdot \vec{\alpha}_j^*$ is a Cartan matrix. The Coulomb brach coordinate for each simple root is equivalent to the on-shell action of each monopole, which is given by 
\begin{align}
V_{\alpha_k} := \exp \left[ \sum_{i=1}^r \vec{\alpha}_k^* \cdot \vec{\alpha}_i^* \phi_i   \right],
\end{align}
where we omitted the normalization of the action. Rigorously speaking, the Coulomb branch operator includes the dual photon which is a dualized scalar from the $U(1)$ photon. Here we omitted it for simplicity since the dual photon dependence is easily restored.  

Since the Coulomb branch coordinates are originally the member of the vector superfield, it is neutral under the flavor symmetries. However, the zero-modes around the monopole background spontaneously break the flavor symmetries. As a result, the Coulomb branch operators have non-trivial charges under the non-linearly realized flavor symmetries \cite{Affleck:1982as} which is the mixing between the original flavor symmetries and the topological $U(1)$ symmetry. The magnitude of the mixing is related to the number of the fermion zero-modes. Hence we need to calculate the zero-modes around the monopole background by employing the Callias index theorem \cite{Callias:1977kg,Weinberg:1979zt,deBoer:1997kr}.

The Callias index theorem claims that the number of fermion zero-modes is obtained by the following formula
\begin{align}
N=\sum_w \frac{1}{2} \mathrm{sign}(w (\phi)) w(g),
\end{align}
where the summation is taken over all the weight of the matters and $g$ is a magnetic charge of the monopole which we consider. $\phi$ is an adjoint scalar field in the vector multiplet.

Let us consider the classical Coulomb branch of a 3d $\mathcal{N}=2$ $Spin(7)$ gauge theory. In our notation, the Weyl chamber which we chose is defined by
\begin{align}
2 \phi_1- \phi_2 &\ge 0\\
-\phi_1+2 \phi_2-2 \phi_3 &\ge 0\\
-\phi_2 +2 \phi_3 &\ge 0.
\end{align}
In order to simplify the Weyl chamber, we sometimes change the variables as
\begin{align}
\phi_1 &=: \sigma_1 \\
\phi_2  &=: \sigma_1 + \sigma_2 \\
\phi_3  &=:\frac{1}{2} (\sigma_1+\sigma_2+\sigma_3).
\end{align}
In this redefinition, the Weyl chamber is simplified to
\begin{align}
\sigma_1 \ge \sigma_2 \ge \sigma_3 \ge 0.
\end{align}
The Coulomb branch operators are defined as
\begin{align}
Y_1 &\simeq \exp  [2 \phi_1 -\phi_2]  =\exp [\sigma_1-\sigma_2]\\
Y_2 &\simeq \exp[- \phi_1 + 2 \phi_2 -2 \phi_3] =\exp [\sigma_2 -\sigma_3]\\
Y_3 & \simeq \exp[-2 \phi_2 +4 \phi_3] = \exp [2 \sigma_3] \\
Z &:= Y_1 Y_2^2 Y_3 \simeq \exp[ \phi_2] =\exp[\sigma_1+\sigma_2] \\
Y &:= \sqrt{Y_1 Z} \simeq \exp [\phi_1] = \exp [\sigma_1],~~~Y_{spin}:=Y_1 Z
\end{align}
where $Z$ corresponds to a lowest co-root and plays an important role when we study the connection between 3d and 4d theories. $Y$ and $Y_{spin}$ were defined in \cite{Aharony:2011ci, Aharony:2013kma}, which are the globally defined Coulomb branch coordinates for the 3d $\mathcal{N}=2$ $O(N)$ or $Spin(N)$ gauge theories with vectorial matters.
By using the Callias index theorem, one can compute the fermion zero-modes around each magnetic monopole. Table \ref{zeromode} summarizes the fermion zero-modes for each operator. Notice that we have to divide the Weyl chamber further into two regions depending on the sign of $\phi_1-\phi_3$ for the spinor zero-modes.

\begin{table}[H]\caption{Fermion zero-modes} 
\begin{center}
  \begin{tabular}{|c||c|c|c| } \hline
  &adjoint&vector&spinor \\ \hline
$Y_1$&2&0&$1+\mathrm{sign(\phi_1-\phi_3)}$\\
$Y_2$&2&0&0 \\
$Y_3$&2&2&$1-\mathrm{sign(\phi_1-\phi_3)}$ \\ \hline
$Z:= Y_1 Y_2^2 Y_3$&8&2&2 \\
$Y:= \sqrt{Y_1 Z}$ $(\phi_1 >\phi_3)$&5&1&2 \\
$Y_{spin}:=Y_1 Z$ $(\phi_1 >\phi_3)$&10&2&4 \\ \hline
  \end{tabular}
  \end{center}\label{zeromode}
\end{table}

For the 3d $\mathcal{N}=2$ pure $Spin(7)$ theory without matters, all the Coulomb branch operators $Y_i$ get two gaugino zero-modes. Thus we have the non-perturbative superpotential like $\frac{1}{Y_i}$ and there is no stable SUSY vacuum. 

When we turn on the matters in a vectorial representation, $Y_3$ gets additional zero-modes from the vectorial fermions and $W=\frac{1}{Y_3}$ is not allowed. As a result, one dimensional Coulomb branch would remain as the (quantum) moduli space. For an $(S)O(7)$ case with vector matters \cite{Aharony:2011ci}, $Y$ is a globally defined one-dimensional Coulomb branch operator. For a $Spin(7)$ theory with vectorial matters the correct coordinate is $Y_{spin}$ \cite{Aharony:2013kma}. In these theories, $Z$ appears when we put  the corresponding 4d theories on a circle. 

Let us next consider the $Spin(7)$ theory with spinorial matters. For $\phi_1 > \phi_3$, $Y_1$ has zero-modes from the spinor in addition to the gaugino zero-modes. Thus, it is expected that $Y_1$ is not lifted and that there is a one-dimensional Coulomb branch for $\phi_1 > \phi_3$. The same argument would be available also for $\phi_1 < \phi_3$ and $Y_3$ is un-lifted. In this theory, we need one globally defined coordinate and we will use $Z$ for parametrizing it.

When both the vectors and the spinors are added into the $Spin(7)$ theory, the Coulomb branch becomes more complicated. For $\phi_1 > \phi_3$, $Y_1$ and $Y_3$ have more than two fermion zero-modes. Hence they are not lifted while $Y_2$ is still lifted via the monopole superpotential. For $\phi_1 < \phi_3$, only $Y_3$ has more than two zero-modes and $Y_{1,2}$ are lifted. We therefore need to introduce two coordinates for the description of the (semi-)classical Coulomb moduli. We expect that one of them would be the operator $Z$.  This is because the zero-mode of $Z$ does not depend on the sign of $\phi_1-\phi_3$ so that $Z$ would be globally defined on the whole Coulomb branch. The other one would be described by $Y$ or $Y_{spin}$. Notice that this analysis is completely (semi-)classical. Therefore the quantum effects might modify these pictures. In fact we will see that the 3d $\mathcal{N}=2$ $Spin(7)$ gauge theories with $N_f$ vectors and $N_S$ spinors sometimes show the one-dimensional Coulomb branch.

\section{3d $\mathcal{N}=2$ $Spin(7)$ theories with spinorial matters}
In a previous section, we studied the (semi-)classical Coulomb branch of the $Spin(7)$ theory. Here we examine the quantum aspects of the $Spin(7)$ Coulomb branch.
Let us start with the 3d $\mathcal{N}=2$ $Spin(7)$ gauge theory with spinorial matters. The Higgs branch is parametrized by a meson $M_{SS}:=SS$ for $N_S \le 3$. The baryonic operator $B_S:=S^4$ is also necessary for $N_S \ge 4$. The matter contents and their quantum numbers are summarized in Table \ref{Tspinor}. The table also includes the dynamical scale $\eta$ of the 4d gauge coupling. Since the $U(1)$ symmetries are anomalous in 4d due to the chiral anomalies, the dynamical scale is charged under the $U(1)$ symmetries. For the Coulomb branch, we predict that $Z$ is a correct monopole operator.

\begin{table}[H]\caption{Quantum numbers for 3d $\mathcal{N}=2$ $Spin(7)$ with $N_f$ spinors} 
\begin{center}
\scalebox{0.88}{
  \begin{tabular}{|c||c||c|c|c| } \hline
  &$Spin(7)$&$SU(N_S)$&$U(1)_S$&$U(1)_R$ \\ \hline
$S$ & $ \mathbf{2^N}=\mathbf{8}$&${\tiny \yng(1)}$&1& $R_S$ \\
$\lambda$ &$\mathbf{Adj.}$&1&0&$1$  \\ \hline
$\eta=\Lambda_{N_f,N_S}^b$&1&1&$2N_S$&$2N_S(R_S-1) +10$  \\ \hline 
$M_{SS}:=SS$&1&$\tiny \yng(2)$&2&$2R_S$ \\
$B_S:=S^4$&1&${\tiny \yng(1,1,1,1)}$&$4$&$4R_S$  \\[10pt]\hline 
$Y_1$&1&1&$-N_S(1+\mathrm{sign}(\phi_1-\phi_3))$& $-2-N_S(R_S-1)(1+\mathrm{sign}(\phi_1-\phi_3))$\\
$Y_2$&1&1&0&$-2$\\
$Y_3$&1&1&$-N_S(1-\mathrm{sign}(\phi_1-\phi_3))$& $-2-N_S(R_S-1)(1-\mathrm{sign}(\phi_1-\phi_3))   $\\ \hline
$Z:=Y_1Y_2^2Y_3$&1&1&$-2N_S$& $-8 -2N_S (R_S-1) $  \\   \hline
  \end{tabular}}
  \end{center}\label{Tspinor}
\end{table}

For any $N_S$, the superpotential $W= \eta Z$ is available, which is dynamically generated from the KK-monopole and necessary when connecting the 3d theory to the 4d theory. 
From Table \ref{Tspinor}, we find that the following superpotentials are consistent with all the symmetries.
\begin{align}
W_{N_S \le 3} &=  \left( \frac{1}{Z \det M_{SS}}  \right)^{\frac{1}{4-N_S}} \\
W_{N_S = 4} &= X  \left[ Z (\det M_{SS} -B_S^2) -1 \right] \\
W_{N_S=5} &= Z  \left(  \det  \, M_{SS} -  B_{S}^i B_{S}^j  M_{SS,ij} \right) 
\end{align}
Consequently, there is no stable SUSY vacuum for $N_S \le 3$. The Higgs and Coulomb branches are quantum-mechanically merged for $N_S=4$. The large values of the Higgs branch is connected to the small value of the Coulomb branch. Importantly the origin of the moduli space is not a vacuum. For $N_S =5$, the theory is s-confining, where the origin belongs to the vacua. For $N_S \ge 6$ we have no simple superpotential. In what follows, we will verify our superpotentials above in various ways. It is easy to check the parity anomaly matching for $N_S=5$. The UV and IR descriptions produce the same anomalies.
By adding the term $\eta Z$ from the KK-monopole, the 4d superpotentials 
\begin{align}
W_{N_S \le 3}^{\mathbb{S}_1 \times \mathbb{R}_3}  &=   \left( \frac{1}{Z \det M_{SS}}  \right)^{\frac{1}{4-N_S}}+\eta Z   \longrightarrow  ~~W_{N_S \le 3}^{4d}= \left( \frac{\eta}{\det \,M_{SS}}  \right)^{\frac{1}{5-N_s}} \\
W_{N_S=4}^{\mathbb{S}_1 \times \mathbb{R}_3} &=  X  \left[ Z (\det M_{SS} -B_S^2) -1 \right] +\eta Z  \longrightarrow ~~ W^{4d}_{N_S=4} = \frac{\eta}{\det \, M_{SS} -B_S^2} \\
W_{N_S=5}^{\mathbb{S}_1 \times \mathbb{R}_3} &=Z  \left(  \det  \, M_{SS} -  B_{S}^i B_{S}^j  M_{SS,ij} \right) +\eta Z  \nonumber \\
& \qquad \longrightarrow ~~ W^{4d}_{N_S=5}= X \left[  \det  \, M_{SS} -  B_{S}^i B_{S}^j  M_{SS,ij} -\eta  \right]
\end{align}
are correctly reproduced.

Next let us  introduce a complex mass deformation. We restrict ourself to the case with $N_S=5$ and introduce a complex mass to the last flavor. By integrating out the massive modes, we arrive at the quantum constraint for $N_S=4$ as follows.
\begin{align}
W=W_{N_S=5} +m M_{SS,55}  \rightarrow  \begin{cases}
    B_S^i=0 & (i=1,2,3,4) \\
    M_{SS,i5} =0 &(i=1,2,3,4)  \\
    m=Z (\det \, \hat{M}_{SS} -B_{S}^5 B_S^{5})
  \end{cases}
\end{align}

We can also test the Higgs branch. When a spinor gets a vev $\braket{M_{SS,N_SN_S}} = v^2$, the theory flows to the 3d $\mathcal{N}=2$ $G_2$ gauge theory with $N_S-1$ fundamentals \cite{Nii:2017npz}. The superpotential above correctly explains this flow. For $N_S=5$, we need the following identification between the $Spin(7)$ and $G_2$ moduli coordinates.
\begin{gather}
M_{SS,ij} =: M^{G_2}_{ij} ~(i,j=1,\cdots,4) \\
B_{S}^i =:vB^i_{G_2}~(i=1,\cdots,4),~~~B_{S}^5 =F_{G_2} 
\end{gather}
The superpotential reduces to
\begin{align}
W=v^2 Z( \det \,M^{G_2} - B^i_{G_2} M_{ij}^{G_2}B^j_{G_2} -F_{G_2}^2 ) =: Z_{G_2}( \det \,M^{G_2} - B^i_{G_2} M_{ij}^{G_2}B^j_{G_2} -F_{G_2}^2 ),
\end{align}
where we absorbed the vev into the monopole operator. This superpotential was first obtained in \cite{Nii:2017npz}. The similar argument can be applied also for $N_S \le 4$ and the $G_2$ superpotentials are reproduced.

Finally we briefly discuss the theory with $N_S \ge 6$. In this case one cannot write down the superpotential. From the analysis of the semi-classical Coulomb branch, it is expected that the Coulomb branch is still one-dimensional (it is labeled by $Z$) and that the quantum moduli space would be identical to the (semi-)classical one. If the fractional power in a superpotential is allowed, one can still write down the ``effective'' superpotential. For $N_S=6$, the superpotential
\begin{align}
W_{N_S=6}=  \left[  Z  \left( \det M_{SS} -M_{SS}^2 B_S^2 -\mathrm{Pf} \, B_S \right) \right]^{\frac{1}{2}}
\end{align}
is consistent with all the symmetries. By adding a term $\eta Z$, the 4d result \eqref{4dF6} is reproduced. However, the fractional power leads to the branch-cut singularities on the origin of the moduli space and we have to introduce new massless degrees of freedom along the singularities. Presumably, some Seiberg dual descriptions would explain these massless modes and a certain superconformal fixed point is realized on the origin of the moduli space. We don't discuss it further in this paper and will tackle with this problem elsewhere.

\subsection{Superconformal Indices}

Since the $Spin(7)$ theory with five spinors exhibits the s-confinement phase, the superconformal index is simple enough and it is computed from the dual side. This would be another check of our analysis. For the definitions of the superconformal indices, see for example \cite{Bhattacharya:2008bja,Kim:2009wb,Imamura:2011su,Imamura:2011uj,Kapustin:2011jm,Spiridonov:2009za,Bashkirov:2011vy,Kim:2013cma}.
 The index on the dual side has the contributions from the meson $M_{SS,ij}$, the baryon $B_S^i$ and the Coulomb branch operator $Z$. We set $R_S =\frac{1}{8}$ for simplicity and use a fugacity $u$ for the global $U(1)_S$ symmetry which rotates the spinor. The full index (or the index of the dual description) becomes

\scriptsize
\begin{align}
 I_{magnetic}^{N_S=5} &=
1+15 u^2 x^{1/4}+125 u^4 \sqrt{x}+\left(\frac{1}{u^{10}}+755 u^6\right) x^{3/4}+\left(3675 u^8+\frac{15}{u^8}\right) x+ \left(15252 u^{10}+\frac{125}{u^6}\right) x^{5/4} \nonumber \\
&\quad +\left(\frac{1}{u^{20}}+55880 u^{12}+\frac{750}{u^4}\right) x^{3/2}+ 5 \left(37004 u^{14}+ \frac{717}{u^2}+\frac{3}{u^{18}}\right) x^{7/4}+\left(562985 u^{16}+\frac{125}{u^{16}}+14402\right) x^2 \nonumber \\
&\quad +\left(\frac{1}{u^{30}}+1594185 u^{18}+\frac{750}{u^{14}}+50245 u^2\right) x^{9/4}+\left(4241879 u^{20}+155550 u^{4}+ \frac{3585}{ u^{12}}+\frac{15}{u^{28}}\right) x^{5/2} \nonumber \\
&\quad +\left(10688125 u^{22}+433550 u^{6}+\frac{14403}{ u^{10}}+\frac{125}{u^{26}}\right) x^{11/4}+\left(\frac{1}{u^{40}}+25661515 u^{24}+\frac{750}{u^{24}}+1097955 u^8+\frac{50270}{u^8}\right) x^3 +\cdots
\end{align}
\normalsize
We will briefly explain the low-lying operators below.
\begin{itemize}
\item
The first term is an identity operator.
\item 
The second term $15 u^2 x^{1/4}$ is identified with a meson contribution $M_{SS,ij}$ which has $15$ independent components.

\item
The third term $125 u^4 \sqrt{x}$ consists of two operators. One is a baryonic operator $B_S^i$ which contributes to the index as $5 u^4 x^{1/2}$ and the other is a square of the mesons $M_{SS} \otimes M_{SS}$, whose flavor indices are symmetrized. Thus we have $\mathbf{15} \times \mathbf{15}|_{symmetric~part} = 120= \overline{\mathbf{50}}+\overline{\mathbf{70}}'$ in an $SU(5)$ notation. 

\item 
The fourth term $\left(\frac{1}{u^{10}}+755 u^6\right) x^{3/4}$ contains the monopole operator which is denoted by $Z$. The remaining parts are the symmetric products of the Higgs branch operators, $M_{SS}^3$ and $M_{SS} B_S$.

\item
The higher order contributions are recognized as composite operators of $M_{SS}, B_S$ and $Z$ by properly symmetrizing the flavor indices.

\end{itemize}

Let us move on to the electric side. The index on the electric side is decomposed into the index for each GNO charge $(m_1,m_2,m_3),~m_i \in \mathbb{Z}/2$. Since we now discuss the $Spin(7)$ gauge group, we have to sum up only the sectors with $m_1+m_2 +m_3 \in \mathbb{Z}$ \cite{Aharony:2013kma}. We need to consider the GNO charges $(0,0,0), \left( \frac{1}{2},\frac{1}{2},0\right),\left( 1,1,0\right),\left( \frac{3}{2},\frac{3}{2},0\right)$ and $(2,2,0)$ up to $O(x^3)$. The index with zero GNO charge becomes

\footnotesize
\begin{align}
I_{electric}^{(0,0,0)} &=1+15 u^2 x^{1/4}+125 u^4 \sqrt{x}+755 u^6 x^{3/4}+3675 u^8 x+15252 u^{10} x^{5/4}+55880 u^{12} x^{3/2} +185020 u^{14} x^{7/4} \nonumber \\
&\quad +\left(562985 u^{16}-25\right) x^2+\left(1594185 u^{18}-400 u^2\right) x^{9/4}  +\left(4241879 u^{20}-3450 u^4\right) x^{5/2}  \nonumber \\
&\qquad +\left(10688125 u^{22}-21200 u^6\right) x^{11/4} +\left(25661515 u^{24}-103775 u^8\right) x^3+\cdots.
\end{align}
\normalsize

\noindent The first term is an identity operator and regarded as the state $\ket{0,0,0}$. Since the gauge group is not broken in this sector, we can freely act the Higgs branch operators on the state $\ket{0,0,0}$. For example, $15 u^2 x^{1/4}$ is identified with $M_{SS,ij} \ket{0,0,0}$. Next let us study the sectors with non-zero GNO charges.

\footnotesize
\begin{align}
I_{eletric}^{\left(\frac{1}{2},\frac{1}{2},0 \right)} &=\frac{x^{3/4}}{u^{10}}+\frac{15 x}{u^8}+\frac{125 x^{5/4}}{u^6}+\frac{750 x^{3/2}}{u^4}+\frac{3585 x^{7/4}}{u^2}+14427 x^2+50645 u^2 x^{9/4} \nonumber  \\
&\qquad +159000 u^4 x^{5/2}+\left(454750 u^6-\frac{24}{u^{10}}\right) x^{11/4}+\frac{5 \left(240346 u^{16}-75\right) x^3}{u^8}+\cdots  \\
I_{electric}^{(1,1,0)}&=\frac{x^{3/2}}{u^{20}}+\frac{15 x^{7/4}}{u^{18}}+\frac{125 x^2}{u^{16}}+\frac{750 x^{9/4}}{u^{14}}+\frac{3585 x^{5/2}}{u^{12}}+\frac{14427 x^{11/4}}{u^{10}}+\frac{50645 x^3}{u^8}+\cdots \\
I_{electric}^{\left(\frac{3}{2},\frac{3}{2},0 \right)} &=\frac{x^{9/4}}{u^{30}}+\frac{15 x^{5/2}}{u^{28}}+\frac{125 x^{11/4}}{u^{26}}+\frac{750 x^3}{u^{24}} + \cdots,~~~~I_{electric}^{(2,2,0)} =\frac{x^3}{u^{40}}+\cdots
\end{align}
\normalsize

\noindent The sector with a GNO charge $\left( \frac{1}{2},\frac{1}{2},0\right)$ contains the monopole operator. The first term $\frac{x^{3/4}}{u^{10}}$ is $Z$ (see Table \ref{Tspinor}) and the corresponding state is expressed as $\ket{\frac{1}{2}, \frac{1}{2},0}$. The proceeding two terms $\frac{15 x}{u^8}+\frac{125 x^{5/4}}{u^6}$ are $M_{SS}\ket{\frac{1}{2}, \frac{1}{2},0}$ and $(M_{SS}^2+B_S)\ket{\frac{1}{2}, \frac{1}{2},0}$ respectively.
The term $\frac{750 x^{3/2}}{u^4}$ needs some explanation. The GNO charge assignment $\left( \frac{1}{2},\frac{1}{2},0\right)$ breaks the gauge group to $ Spin(3) \times SU(2) \times U(1)$. The spinor reduces to $(\mathbf{2},\mathbf{2})_0$ where we omitted the charged fields since we cannot act the charged fields on $\ket{\frac{1}{2}, \frac{1}{2},0}$ a la \cite{Bashkirov:2011vy}. Therefore we cannot totally anti-symmetrize the $SU(5)$ flavor indices of the reduced spinors (fourth-order anti-symmetrization is still allowed). Therefore, in the product $M_{SS} \times B_S = \mathbf{15} \otimes  \bar{\mathbf{5}} = \mathbf{5} +\mathbf{70}  $,  we have to discard the $\mathbf{5}$ representation. 
As a result, $\frac{750 x^{3/2}}{u^4}$ is regarded as $(M_{SS}^3 +M_{SS}B_S)\ket{\frac{1}{2}, \frac{1}{2},0}$.
The similar argument is available for higher order terms. By summing up these indices, we observe the complete agreement between the electric and magnetic sides.

\section{3d $\mathcal{N}=2$ $Spin(7)$ with vector and spinor matters}
Let us next study the 3d $\mathcal{N}=2$ $Spin(7)$ gauge theory with $N_f$ vector matters and with $N_S$ spinor maters. From the analysis in Section 3, one might expect that the quantum Coulomb branch is two-dimensional. However the previous argument was semi-classical and it would be generally modified. In fact, we will see that the dimension of the Coulomb branch extremely depends on the matter contents. In this section, we are mostly interested in s-confinement phases. We will find the s-confining descriptions for $(N_f,N_S)=(0,5),(1,4),(2,3),(3,2)$ and $(4,1)$. Since we have already discussed the $(N_f,N_S)=(0,5)$ case, we start with $(N_f, N_S) =(1,4) $. The dynamics for the theories with fewer matters can be obtained from the s-confinement description by integrating out massive fields.

\subsection{$(N_f, N_S) =(1,4) $}
From the (semi-)classical analysis of the Coulomb branch operators for the simple roots, the Coulomb moduli should be divided into two parts depending on the sign of $\phi_1 -\phi_3$. Thus we expected that two (quantum) Coulomb moduli $Z$ and $Y$(or $Z$ and $Y_{spin}$) are necessary. However, in this phase with $(N_f,N_S)=(1,4)$, we can relate these two coordinates by acting the Higgs branch coordinates on the monopole operator. For example, $Y$ has the same quantum numbers as $Z^2 \times (\alpha B'_S P_{A,1}^2 +\beta M_{QQ}B_SB'_S)$ with some numerical coefficients $\alpha$ and $\beta$. The deep reason behind this identification is unclear, but it is allowed at least from a symmetry argument. We predict that the (quantum) Coulomb branch is described by a single $Z$ coordinate. The validity of this prediction can be checked via various deformations and the superconformal indices below. 

For the description of the Higgs branch, we define following operators.
\begin{gather}
M_{QQ}:=QQ,~~~ M_{SS} :=S^2 \\
P_{A1}:= SQS,~~~B_{S} := S^4 ,~~~B'_{S} := S^4Q 
\end{gather}
We listed the quantum numbers of the matter contents and of the moduli coordinates in Table \ref{T14}.
From the table, one can write down the superpotential
\begin{align}
W=Z \left[ M_{QQ} (\det \, M_{SS}- B_S^2 ) +B'^2_{S} +B_S P_{A1}^2 +M_{SS}^2 P_{A1}^2  \right]  +\eta Z,
\end{align}
where the last term is generated by a KK-monopole and absent in a 3d limit. By integrating out the Coulomb branch, we obtain a quantum constraint in 4d. This IR description gives the same parity anomalies as the UV theory.  In addition, we cannot satisfy the parity anomaly matching if we introduce two Coulomb branch operators. This is a first non-trivial check of our prediction.

\begin{table}[H]\caption{Quantum numbers for $(N_f,N_S)=(1,4)$} 
\begin{center}
\scalebox{0.76}{
  \begin{tabular}{|c||c||c|c|c|c| } \hline
  &$Spin(7)$&$SU(4)$&$U(1)_Q$&$U(1)_S$&$U(1)_R$ \\ \hline
Q& ${\tiny \yng(1)}$&1&1&0& $R_f$ \\
$S$ & $\mathbf{2^{N}}=\mathbf{8}$&${\tiny \yng(1)}$&0&1& $R_S$ \\
$\lambda$ &$\mathbf{Adj.}$&1&0&0&$1$  \\ \hline
$\eta=\Lambda_{N_f,N_S}^b$&1&1&$2$&$8$&$2(R_f-1)+8(R_S-1) +10=2R_f+8R_S$  \\ \hline 
$M_{QQ}:=QQ$&1&1&2&0&$2R_f$ \\
$M_{SS}:=SS$&1&$\tiny \yng(2)$&0&2&$2R_S$ \\
$B_S:=S^4$&1&1&0&$4$&$4R_S$  \\
$B'_S:=S^4 Q$&1&1&1&4& $R_f+4R_S$ \\
$P_{A1}:=SQS$&1&${\tiny \yng(1,1)}$&1&2&$R_f+2R_S$ \\ \hline
$Z:=Y_1Y_2^2Y_3$&1&1&$-2$&$-8$& $-8 -8(R_S-1) -2(R_f-1)=2-2R_f-8R_S$  \\ 
$Y:=\sqrt{Y_1 Z}$ $(\phi_1 \ge \phi_3)$ &1&1&$-1$&$-8$& $-5-(R_f-1) -8(R_S-1)=4-R_f-8R_S$  \\
$Y_{spin}:=Y_1^2Y_2^2Y_3$ $(\phi_1 \ge \phi_3)$&1&1&$-2$&$-16$&$-10-2(R_f-1)-16(R_S-1)=8-2R_f-16R_S$ \\ \hline
  \end{tabular}}
  \end{center}\label{T14}
\end{table}

In order to test the superpotential above, let us consider various directions of the Higgs branch, which would justify our analysis. First, we consider introducing the vectorial vev $\braket{M_{QQ}} =v^2 $ which breaks the $Spin(7)$ group to $Spin(6) \cong SU(4)$. The low-energy theory is a 3d $\mathcal{N}=2$ $SU(4)$ gauge theory with four flavors in a (anti-)fundamental representation, which is s-confining \cite{Aharony:1997bx}. Since the global symmetries are enhanced to $SU(4) \times SU(4)$ in the low-energy limit, we have to rename and decompose the fields as
\begin{gather}
M_{SS,ij} =: M_i^j+M_j^i ~~~(i,j=1,\cdots,4)\\
B_{S}=:\frac{v}{4}(B+\bar{B}),~~~B'_S =:\frac{v}{4}(B-\bar{B}) \\
P_{A1}=:v (M_i^j -M_j^i),
\end{gather}
where $ M_i^j$ is regarded as a meson and $B,\bar{B}$ are (anti-)baryonic operators in the $SU(4)$ theory. By absorbing the vev into the redifinition of the monopole operator, the superpotential reduces to
\begin{align}
W= v^2 Z \left[ \det M-B\bar{B} \right] =:Y_{SU(4)}  \left[ \det M-B\bar{B} \right],
\end{align}
which is precisely the superpotential of the 3d $\mathcal{N}=2$ $SU(4)$ theory with four flavors \cite{Aharony:1997bx}.

Next, let us focus on the $G_2$ direction of the Higgs branch, which is achieved by introducing a vev for a single spinorial field as $\braket{M_{SS,44}}=v^2$. The low-energy theory becomes a 3d $\mathcal{N}=2$ $G_2$ gauge theory with four fundamental matters, which is again s-confining \cite{Nii:2017npz}. Although the vev breaks the global $SU(4)$ symmetry to $SU(3)$, we again have the enhanced $SU(4)$ symmetry at the low-energy limit since the vector and the spinors become the same representation in $G_2$. We need the following identification between the $Spin(7)$ and $G_2$ moduli coordinates.
\begin{gather}
M_{SS,ij} =: M_{ij}^{G_2} ~~(i,j=1,2,3),~~~~
M_{QQ}=: M_{44}^{G2}\\
B_{S}=:vB_{G_2}^4,~~~B'_S =:v F \\
\epsilon^{ijk}P_{A1, jk}=: B^i_{G_2},~~~~P_{A1,i4}=: vM_{i,4}^{G_2}~~(i,j,k=1,2,3) 
\end{gather}
The superpotential reduces to 
\begin{align}
W=v^2 Z \left[ \det M^{G_2} +F^2 +B^a M_{ab} B^c \right]:=Z_{G_2} \left[ \det M^{G_2} +F^2 +B^a M_{ab} B^c \right],
\end{align}
which is the superpotential observed in \cite{Nii:2017npz}. 

We can also consider a complex mass deformation for a vectorial matter.  By introducing the mass term $m M_{QQ}$, we find that $B'_S$ and $P_{A,1}$ are integrated out. The equation of motion for $M_{QQ}$ leads to a quantum constraint 
\begin{align}
m+Z(\det \, M_{SS} -B_{S}^2)=0,
\end{align}
 which was observed in a previous section with $N_S=4$.

 \subsubsection*{Superconformal Indices}
As an additional test of our analysis, we compute the superconformal index of the 3d $\mathcal{N}=2$ $Spin(7)$ theory with $(N_f, N_S) =(1,4) $. Since the theory is s-confining, the dual description does not contain any gauge group. The index on the dual side is

\scriptsize
\begin{align}
I_{dual} &=1+x^{1/4} \left(t^2+10 u^2\right)+6 t u^2 x^{3/8}+\sqrt{x} \left(t^4+10 t^2 u^2+56 u^4\right)+x^{5/8} \left(6 t^3 u^2+61 t u^4\right) \nonumber \\
&\quad +x^{3/4} \left(t^6+10 t^4 u^2+\frac{1}{t^2 u^8}+77 t^2 u^4+230 u^6\right)+x^{7/8} \left(6 t^5 u^2+61 t^3 u^4+346 t u^6\right) \nonumber \\
&\quad+x \left(t^8+10 t^6 u^2+77 t^4 u^4+446 t^2 u^6+\frac{10}{t^2 u^6}+771 u^8+\frac{1}{u^8}\right)+x^{9/8} \left(6 t^7 u^2+61 t^5 u^4+402 t^3 u^6+1436 t u^8+\frac{6}{t u^6}\right) \nonumber \\
&\quad +x^{5/4} \left(t^{10}+10 t^8 u^2+77 t^6 u^4+446 t^4 u^6+2007 t^2 u^8+\frac{t^2}{u^8}+\frac{56}{t^2 u^4}+2232 u^{10}+\frac{10}{u^6}\right) \nonumber \\
&\quad +x^{11/8} \left(6 t^9 u^2+61 t^7 u^4+402 t^5 u^6+2017 t^3 u^8+4856 t u^{10}+\frac{6 t}{u^6}+\frac{60}{t u^4}\right) \nonumber \\
&\quad +x^{3/2} \left(t^{12}+10 t^{10} u^2+77 t^8 u^4+446 t^6 u^6+\frac{1}{t^4 u^{16}}+2133 t^4 u^8+\frac{t^4}{u^8}+7398 t^2 u^{10}+\frac{10 t^2}{u^6}+\frac{230}{t^2 u^2}+5776 u^{12}+\frac{76}{u^4}\right) \nonumber \\
&\quad +\cdots,
\end{align}
\normalsize

\noindent where we set $R_f=R_S=\frac{1}{8}$ for simplicity. $t$ and $u$ are the fugacities for the $U(1)_Q$ and $U(1)_S$ symmetries respectively. The magnetic index has the contributions from $M_{QQ},M_{SS},B_S,B'_S,P_{A1}$ and $Z$.

For the index on the electric side, we have to sum up the indices from the GNO charges $(0,0,0),\left( \frac{1}{2},\frac{1}{2},0 \right)$ and $(1,1,0)$ up to $O(x^{3/2})$. Remember that the GNO charge $(m_1,m_2,m_3)$ must satisfy the relation $\sum m_i \in \mathbb{Z}$. The electric indices are

\tiny
\begin{align}
I_{electric}^{(0,0,0)} &=1+x^{1/4} \left(t^2+10 u^2\right)+6 t u^2 x^{3/8}+\sqrt{x} \left(t^4+10 t^2 u^2+56 u^4\right)+x^{5/8} \left(6 t^3 u^2+61 t u^4\right) \nonumber \\
&\qquad +x^{3/4} \left(t^6+10 t^4 u^2+77 t^2 u^4+230 u^6\right)+x^{7/8} \left(6 t^5 u^2+61 t^3 u^4+346 t u^6\right) \nonumber \\
&\qquad +x \left(t^8+10 t^6 u^2+77 t^4 u^4+446 t^2 u^6+771 u^8\right)+x^{9/8} \left(6 t^7 u^2+61 t^5 u^4+402 t^3 u^6+1436 t u^8\right) \nonumber \\
& \qquad +x^{5/4} \left(t^{10}+10 t^8 u^2+77 t^6 u^4+446 t^4 u^6+2007 t^2 u^8+2232 u^{10}\right)\nonumber \\
&\qquad +x^{11/8} \left(6 t^9 u^2+61 t^7 u^4+402 t^5 u^6+2017 t^3 u^8+4856 t u^{10}\right) \nonumber \\
&\qquad+x^{3/2} \left(t^{12}+10 t^{10} u^2+77 t^8 u^4+446 t^6 u^6+2133 t^4 u^8+7398 t^2 u^{10}+5776 u^{12}\right)+\cdots \\
I^{\left( \frac{1}{2},\frac{1}{2},0 \right)}_{electric} &=\frac{x^{3/4}}{t^2 u^8}+x \left(\frac{10}{t^2 u^6}+\frac{1}{u^8}\right)+\frac{6 x^{9/8}}{t u^6}+x^{5/4} \left(\frac{t^2}{u^8}+\frac{56}{t^2 u^4}+\frac{10}{u^6}\right)+x^{11/8} \left(\frac{6 t}{u^6}+\frac{60}{t u^4}\right)+x^{3/2} \left(\frac{t^4}{u^8}+\frac{10 t^2}{u^6}+\frac{230}{t^2 u^2}+\frac{76}{u^4}\right)+\cdots \\
I_{electric}^{(1,1,0)} &=\frac{x^{3/2}}{t^4 u^{16}} +\cdots
\end{align}
\normalsize

\noindent From the sector with zero GNO charge, we can read off the Higgs branch operators. The second term $x^{1/4} \left(t^2+10 u^2\right)$ represents the mesonic operators $M_{QQ}$ and $M_{SS,ij}$. The third term $6 t u^2 x^{3/8}$ corresponds to $P_{A1,ij}$. The baryonic operators $B_S$ and $B'_S$ are represented as $u^4 x^{1/2}$ and $tu^4 x^{5/8}$ respectively. The higher order terms are the products of the Higgs branch operators, whose flavor indices have to be symmetrized. The index with a GNO charge $\left( \frac{1}{2},\frac{1}{2},0 \right)$ contains the monopole operator. The first term $\frac{x^{3/4}}{t^2 u^8}$ can be regarded as $Z$ (see Table \ref{T14}). The index with a GNO charge $(1,1,0)$ represents $Z^2$. By summing up these three sectors we observe exact matching between the magnetic and electric indices.

\subsection{$(N_f, N_S) =(2,3) $}
Let us next consider the 3d $\mathcal{N}=2$ $Spin(7)$ gauge theory with two vectors and three spinors (see Table \ref{T23}).  
In this case, we also have a similar relation between $Z^3$ and $Y_{spin}$. Therefore, we expect that the quantum Coulomb branch is one-dimensional although the (semi-)classical analysis suggested the two-dimensional coordinates. We use the coordinate $Z$ to parametrize the Coulomb branch. The Higgs branch is described by the following operators
\begin{gather}
M_{QQ}:=QQ,~~~M_{SS}:=S^2 \\
P_{A1}:= SQS,~~~ P_{A2}:=SQ^2S.
\end{gather}
Notice that the spinors and the vectors now can be anti-symmetrized and we omitted the gamma matrices above for simplicity. The superpotential consistent with all the symmetries is 
\begin{align}
W=Z   \left[ \det \, M_{QQ} \det \, M_{SS} +P_{A1}^2 M_{QQ} M_{SS}+P_{A1}^2 P_{A2} +P_{A2}^2 M_{SS}   \right] +\eta Z, \label{W23}
\end{align}
where the last term exists only when we put the theory on $\mathbb{S}^1 \times \mathbb{R}^3$. By integrating out the Coulomb branch operator, we obtain a 4d quantum constraint.

\begin{table}[H]\caption{Quantum numbers for $(N_f, N_S) =(2,3) $} 
\begin{center}
\scalebox{0.71}{
  \begin{tabular}{|c||c||c|c|c|c|c| } \hline
  &$Spin(7)$&$SU(2)$&$SU(3)$&$U(1)_Q$&$U(1)_S$&$U(1)_R$ \\ \hline
Q& ${\tiny \yng(1)}$&${\tiny \yng(1)}$&1&1&0& $R_f$ \\
$S$ & $\mathbf{2^{N}}=\mathbf{8}$&1&${\tiny \yng(1)}$&0&1& $R_S$ \\
$\lambda$ &$\mathbf{Adj.}$&1&1&0&0&$1$  \\ \hline
$\eta=\Lambda_{N_f,N_S}^b$&1&1&1&$4$&$6$&$4(R_f-1)+6(R_S-1) +10$  \\ \hline 
$M_{QQ}:=QQ$&1&$\tiny \yng(2)$&1&2&0&$2R_f$ \\
$M_{SS}:=SS$&1&1&$\tiny \yng(2)$&0&2&$2R_S$ \\
$P_{A1}:=SQS$&1&$\tiny \yng(1)$&$\tiny \overline{\yng(1)}$&1&2&$R_f+2R_Q$ \\
$P_{A2}:=SQ^2S$&1&1&$\tiny \overline{\yng(1)}$&2&2&$2R_f+2R_S$ \\ \hline
$Z:=Y_1Y_2^2Y_3$&1&1&1&$-4$&$-6$& $-8  -4 (R_f-1)-6 (R_S-1)=2-4R_f-6R_S$  \\ 
$Y_{spin}:=Y_1^2Y_2^2Y_3$ $(\phi_1 \ge \phi_3)$&1&1&1&$-4$&$-12$&$-10-4(R_f-1)-12(R_S-1)=6-4R_f-12R_S$ \\ \hline
  \end{tabular}}
  \end{center}\label{T23}
\end{table}

Let us confirm the validity of the superpotential \eqref{W23}. The UV and IR descriptions yield the same parity anomalies. 
As in the previous case, we can test the $SU(4)$ Higgs branch with $\braket{M_{QQ,22}} =v^2$, where the theory reduces to a 3d $\mathcal{N}=2$ $SU(4)$ gauge theory with one antisymmetric matter and with three (anti-)fundamental flavors. It is not known in the literature whether this low-energy theory is s-confining or not. However we can show that this theory indeed exhibits an s-confinement phase. Table \ref{Tsu41anti} shows the matter contents and their quantum numbers of the $SU(4)$ theory.  
\begin{table}[H]\caption{Quantum numbers for $SU(4)$ with ${\tiny \protect\yng(1,1)}$ and 3 $({\tiny \protect\yng(1)}+{\tiny \overline{\protect\yng(1)}})$} 
\begin{center}
\scalebox{1}{
  \begin{tabular}{|c||c||c|c|c|c|c|c| } \hline
  &$SU(4)$&$SU(3)$&$SU(3)$&U(1)&$U(1)$&$U(1)$&$U(1)_R$ \\ \hline
 $A$ &${\tiny \yng(1,1)}$&1&1&1&0&0&$R_A$ \\
  $Q$&{\tiny \yng(1)}&${\tiny \yng(1)}$&1&0&1&1&$R_Q$ \\
  $\tilde{Q}$&${\tiny \overline{\yng(1)}}$&1&${\tiny \yng(1)}$&0&$-1$&1&$R_Q$ \\ \hline 
$T:=A^2$  &1&1&1&2&0&0&$2R_A$ \\
  $M:=Q\tilde{Q}$&1&${\tiny \yng(1)}$&${\tiny \yng(1)}$&0&0&2&$2R_Q$ \\
   $B_A:=AQ^2$ &1&${\tiny \overline{ \yng(1)}}$&1&1&2&2&$R_A+2R_Q$ \\
 $\bar{B}_A:=A \tilde{Q}^2$ &1&1&${\tiny \overline{\yng(1)}}$&1&$-2$&2&$R_A+2R_Q$  \\ \hline
$Y_{SU(4)}$ &1&1&1&$-2$&0&$-6$&$2-2R_A-6R_Q$ \\ \hline
  \end{tabular}}
  \end{center}\label{Tsu41anti}
\end{table}
The Coulomb branch $Y_{SU(4)}$ corresponds to the breaking $SU(4) \rightarrow SU(2) \times U(1) \times U(1)$. The non-perturbative superpotential becomes
\begin{align}
W=Y_{SU(4)} (T \det \, M+M B_A \bar{B}_A). \label{su41anti}
\end{align}
In deriving the above, we assumed that the Coulomb branch is one-dimensional. This is plausible because  the theory flows to a theory with one-dimensional Coulomb branch along the Higgs branch. For instance, when $M$ gets a vev with rank 1, the low-energy theory becomes a 3d $\mathcal{N}=2$ $SU(3)$ gauge theory with three (anti-)fundamental flavors. This theory has one Coulomb branch coordinate and is also s-confining \cite{Aharony:1997bx}. When $B_A$ or $\bar{B}_A$ gets an expectation value, the theory flows to a 3d $\mathcal{N}=2$ $SU(2)$ with four fundamental matters, which is again s-confining and has a one-dimensional Coulomb branch. Finally, when $T$ gets a vev, the theory flows to a 3d $\mathcal{N}=2$ $USp(4)$ theory with six fundamentals, which is s-confining and has a one-dimensional Coulomb branch.

We can derive the superpotential \eqref{su41anti} from \eqref{W23}. Since the global symmetry is enhanced to $SU(3) \times SU(3)$, we decompose the Higgs branch operators as 
\begin{gather}
M_{QQ,11} =: T,~~~M_{SS,ij}=M_{i \underline{j}} +M_{j \underline{i}} \\
P_{A,1} = \left(
    \begin{array}{c}
      B_A^i+\bar{B}_A^{\underline{i}} \\
      v \epsilon^{ijk} (M_{j \underline{k}} - M_{k \underline{j}} )
    \end{array}
  \right),~~~P_{A,2}^i = v (B_A^i -\bar{B}_A^{\underline{i}}).
\end{gather}
By properly rescaling the Coulomb branch operator $Z$, we arrive at the $SU(4)$ superpotential \eqref{su41anti}.

We can also test the $G_2$ Higgs branch $ \braket{M_{SS,33}}=v^2$, where the theory reduces to a 3d $\mathcal{N}=2$ $G_2$ gauge theory with four fundamental matters, which is s-confining. We can derive the matter contents and the superpotential of the $G_2$ theory from our superpotential. We have to decompose the fields as follows.
\begin{gather}
M_{QQ,ij}=M_{ij}^{G_2}~~(i,j=1,2),~~~~M_{SS,ab}=M_{a+2,b+2}^{G2}~~(a,b=1,2) \\
P_{A1} = \begin{pmatrix}
vM_{14}^{G_2} & -vM_{13}^{G_2} & B^2_{G_2}\\
vM_{24}^{G_2} & -vM_{23}^{G_2} &  -B_{G_2}^1\\
\end{pmatrix},~~~P_{A2}=(v B_{G_2}^3 ,v B_{G_2}^4 ,F_{G_2})
\end{gather}
By substituting these expressions into the superpotential \eqref{W23}, the $G_2$ superpotential is reproduced although unnecessary terms like $F_{G_2}(M^{G_2}_{13}M^{G_2}_{24}-M^{G_2}_{14}M^{G_2}_{23})$ are also generated. Presumably, this is because our description only respects $SU(2) \times SU(2) \times U(1) \subset SU(4)$ of the $G_2$ theory. In the RG flow, these terms are supposed to be suppressed.

\subsubsection*{Superconformal Indices of $SU(4)$ with ${\tiny \protect\yng(1,1)}$ and 3 $({\tiny \protect\yng(1)}+{\tiny \overline{\protect\yng(1)}})$}
We start with the index of the 3d $\mathcal{N}=2$ $SU(4)$ gauge theory with one anti-symmetric matter and with three (anti-)fundamental flavors. Since the theory is s-confining, the confined description also yields the same index. The dual index has the contributions from $T,M,B_A,\bar{B}_A$ and $Y_{SU(4)}$. The dual index becomes
 
 \tiny
\begin{align}
I_{dual}^{SU(4)} &=1+x^{1/3} \left(9 t^2+u^2\right)+6 t^2 u \sqrt{x}+x^{2/3} \left(\frac{1}{t^6 u^2}+45 t^4+9 t^2 u^2+u^4\right)+6 t^2 u x^{5/6} \left(9 t^2+u^2\right) \nonumber \\
&\qquad +x \left(165 t^6+\frac{1}{t^6}+66 t^4 u^2+\frac{9}{t^4 u^2}+9 t^2 u^4+u^6\right)+x^{7/6} \left(\frac{6}{t^4 u}+6 t^2 u \left(45 t^4+9 t^2 u^2+u^4\right)\right) \nonumber \\
& \qquad +x^{4/3} \left(\frac{1}{t^{12} u^4}+495 t^8+353 t^6 u^2+\frac{u^2}{t^6}+66 t^4 u^4+\frac{9}{t^4}+9 t^2 u^6+\frac{45}{t^2 u^2}+u^8\right) \nonumber \\
&\qquad +x^{3/2} \left(990 t^8 u+326 t^6 u^3+54 t^4 u^5+\frac{6 u}{t^4}+6 t^2 u^7+\frac{48}{t^2 u}\right) \nonumber \\
&\qquad +x^{5/3} \left(\frac{1}{t^{12} u^2}+\frac{9}{t^{10} u^4}+1287 t^{10}+1431 t^8 u^2+353 t^6 u^4+\frac{u^4}{t^6}+66 t^4 u^6+\frac{9 u^2}{t^4}+9 t^2 u^8+\frac{57}{t^2}+u^{10}+\frac{164}{u^2}\right)+ \cdots,
\end{align}

\normalsize
\noindent where we set $R_A=R_Q=\frac{1}{6}$ for simplicity. $t$ is a fugacity for the $U(1)$ axial symmetry and $u$ counts the number of the anti-symmetric tensor. We did not include the fugacity for $U(1)$ baryon symmetry.

For the electric side, we have to sum up the following sectors up to $O(x^{5/3})$.

\tiny

\begin{align}
I^{(0,0,0)}_{electric}&= 1+x^{1/3} \left(9 t^2+u^2\right)+6 t^2 u \sqrt{x}+x^{2/3} \left(45 t^4+9 t^2 u^2+u^4\right)+x^{5/6} \left(54 t^4 u+6 t^2 u^3\right)+x \left(165 t^6+66 t^4 u^2+9 t^2 u^4+u^6\right)\nonumber \\
&\qquad +x^{7/6} \left(270 t^6 u+54 t^4 u^3+6 t^2 u^5\right)+x^{4/3} \left(495 t^8+353 t^6 u^2+66 t^4 u^4+9 t^2 u^6+u^8\right) \nonumber \\
&\qquad +x^{3/2} \left(990 t^8 u+326 t^6 u^3+54 t^4 u^5+6 t^2 u^7\right)+x^{5/3} \left(1287 t^{10}+1431 t^8 u^2+353 t^6 u^4+66 t^4 u^6+9 t^2 u^8+u^{10}\right)+\cdots \\
I^{\left(\frac{1}{2},0,0 \right)}_{electric}&=\frac{x^{2/3}}{t^6 u^2}+x \left(\frac{1}{t^6}+\frac{9}{t^4 u^2}\right)+\frac{6 x^{7/6}}{t^4 u}+x^{4/3} \left(\frac{u^2}{t^6}+\frac{9}{t^4}+\frac{45}{t^2 u^2}\right)+x^{3/2} \left(\frac{6 u}{t^4}+\frac{48}{t^2 u}\right)+x^{5/3} \left(\frac{u^4}{t^6}+\frac{9 u^2}{t^4}+\frac{57}{t^2}+\frac{164}{u^2}\right)+\cdots \\
I^{(1,0,0)}_{electric} &=\frac{x^{4/3}}{t^{12} u^4}+\frac{x^{5/3} \left(9 t^2+u^2\right)}{t^{12} u^4}+\cdots
\end{align}

\normalsize

\noindent The sector with zero GNO charge contains the Higgs branch operators. The second term $x^{1/3} \left(9 t^2+u^2\right)$ corresponds to $M$ and $T$. The third term $6 t^2 u \sqrt{x}$ is the baryonic operators $B_{A}$ and $\bar{B}_A$. The sector with a GNO charge $\left(\frac{1}{2},0,0 \right)$ contains the Coulomb branch operator $Y_{SU(4)}$. We observe exact matching of the indices between the electric and magnetic sides.

\subsubsection*{Superconformal Indices of $Spin(7)$ with $(N_f,N_S)=(2,3)$}

Let us also examine the superconformal indices of the 3d $\mathcal{N}=2$ $Spin(7)$ gauge theory with $(N_f,N_S)=(2,3)$, which is s-confining. First, the magnetic index has the contributions from $M_{QQ},M_{SS},P_{A1},P_{A2}$ and $Z$. The index can be expanded as

\tiny

\begin{align}
I_{magnetic}&=1+x^{1/4} \left(3 t^2+6 u^2\right)+6 t u^2 x^{3/8}+\sqrt{x} \left(6 t^4+21 t^2 u^2+21 u^4\right)+x^{5/8} \left(18 t^3 u^2+36 t u^4\right) \nonumber \\
&+x^{3/4} \left(10 t^6+\frac{1}{t^4 u^6}+45 t^4 u^2+102 t^2 u^4+56 u^6\right)+x^{7/8} \left(36 t^5 u^2+126 t^3 u^4+126 t u^6\right) \nonumber \\
&+x \left(15 t^8+78 t^6 u^2+249 t^4 u^4+\frac{6}{t^4 u^4}+357 t^2 u^6+\frac{3}{t^2 u^6}+126 u^8\right) \nonumber \\
&+x^{9/8} \left(60 t^7 u^2+270 t^5 u^4+542 t^3 u^6+\frac{6}{t^3 u^4}+336 t u^8\right) \nonumber \\
&+x^{5/4} \left(21 t^{10}+120 t^8 u^2+462 t^6 u^4+1001 t^4 u^6+\frac{21}{t^4 u^2}+987 t^2 u^8+\frac{21}{t^2 u^4}+252 u^{10}+\frac{6}{u^6}\right) \nonumber \\
&+x^{11/8} \left(90 t^9 u^2+468 t^7 u^4+1284 t^5 u^6+1722 t^3 u^8+\frac{36}{t^3 u^2}+756 t u^{10}+\frac{18}{t u^4}\right) \nonumber \\
&+x^{3/2} \left(28 t^{12}+171 t^{10} u^2+\frac{1}{t^8 u^{12}}+741 t^8 u^4+1998 t^6 u^6+3207 t^4 u^8+\frac{56}{t^4}+2310 t^2 u^{10}+\frac{10 t^2}{u^6}+\frac{99}{t^2 u^2}+462 u^{12}+\frac{45}{u^4}\right) +\cdots, \label{Imag23}
\end{align}

\normalsize

\noindent  where $t$ and $u$ are the fugacities for the $U(1)_Q$ and $U(1)_S$ symmetries. We set $R_f=R_S=\frac{1}{8}$ for simplicity.

For the electric side, the index is decomposed into the sectors with different GNO charges. We have to sum up the following sectors up to $O(x^{3/2})$.

\scriptsize

\begin{align}
I_{electric}^{(0,0,0)} &=1+x^{1/4} \left(3 t^2+6 u^2\right)+6 t u^2 x^{3/8}+\sqrt{x} \left(6 t^4+21 t^2 u^2+21 u^4\right)+x^{5/8} \left(18 t^3 u^2+36 t u^4\right) \nonumber \\& \qquad +x^{3/4} \left(10 t^6+45 t^4 u^2+102 t^2 u^4+56 u^6\right)+x^{7/8} \left(36 t^5 u^2+126 t^3 u^4+126 t u^6\right) \nonumber \\
&\qquad +x \left(15 t^8+78 t^6 u^2+249 t^4 u^4+357 t^2 u^6+126 u^8\right) +2 x^{9/8} \left(30 t^7 u^2+135 t^5 u^4+271 t^3 u^6+168 t u^8\right) \nonumber  \\
&\qquad +x^{5/4} \left(21 t^{10}+120 t^8 u^2+462 t^6 u^4+1001 t^4 u^6+987 t^2 u^8+252 u^{10}\right)\nonumber \\
&\qquad+x^{11/8} \left(90 t^9 u^2+468 t^7 u^4+1284 t^5 u^6+1722 t^3 u^8+756 t u^{10}\right) \nonumber \\
&\qquad +x^{3/2} \left(28 t^{12}+171 t^{10} u^2+741 t^8 u^4+1998 t^6 u^6+3207 t^4 u^8+2310 t^2 u^{10}+462 u^{12}\right)+\cdots \\
I_{electric}^{\left(\frac{1}{2},\frac{1}{2},0 \right)} &= \frac{x^{3/4}}{t^4 u^6}+x \left(\frac{6}{t^4 u^4}+\frac{3}{t^2 u^6}\right)+\frac{6 x^{9/8}}{t^3 u^4}+x^{5/4} \left(\frac{21}{t^4 u^2}+\frac{21}{t^2 u^4}+\frac{6}{u^6}\right)+x^{11/8} \left(\frac{36}{t^3 u^2}+\frac{18}{t u^4}\right) \nonumber \\
&\qquad +x^{3/2} \left(\frac{56}{t^4}+\frac{10 t^2}{u^6}+\frac{99}{t^2 u^2}+\frac{45}{u^4}\right) +\cdots \\
I_{electric}^{(1,1,0)} &=  \frac{x^{3/2}}{t^8 u^{12}} +\cdots
\end{align}
\normalsize

\noindent The sector with zero GNO charge contains the Higgs branch operators. The second term $x^{1/4} \left(3 t^2+6 u^2\right)$ corresponds to the mesons $M_{QQ}$ and $M_{SS}$. The third term $6 t u^2 x^{3/8}$ represents $P_{A1}$ while $P_{A2}$ appears as $3 t^2 u^2 x^{1/2}$. The monopole operator is contained in the sector with a GNO charge $\left(\frac{1}{2},\frac{1}{2},0 \right)$. The first term $\frac{x^{3/4}}{t^4 u^6}$ is identified with the monopole operator $Z$. The proceeding terms are regarded as the products between $Z$ and the Higgs branch operators. By summing up these three sectors, we reproduce the magnetic index \eqref{Imag23}.

\subsection{$(N_f,N_S) =(3,2) $}
Let us move on to the 3d $\mathcal{N}=2$ $Spin(7)$ gauge theory with three vectors and two spinors. This case will require two Coulomb branch coordinates even at a quantum level. First, we enumerate the Higgs branch coordinates.
\begin{gather}
M_{QQ} :=QQ,~~~M_{SS} :=S^2\\
P_{S3} := SQ^3S,~~~P_{A1}:=SQS,~~~ P_{A2} :=SQ^2S  
\end{gather}
Table \ref{T32} below shows the matter contents and their quantum numbers. We also listed the 4d dynamical scale and the moduli coordinates.
\begin{table}[H]\caption{Quantum numbers for $(N_f,N_S) =(3,2) $} 
\begin{center}
\scalebox{0.75}{
  \begin{tabular}{|c||c||c|c|c|c|c| } \hline
  &$Spin(7)$&$SU(3)$&$SU(2)$&$U(1)_Q$&$U(1)_S$&$U(1)_R$ \\ \hline
Q& ${\tiny \yng(1)}$&${\tiny \yng(1)}$&1&1&0& $R_f$ \\
$S$ & $\mathbf{2^{N}}=\mathbf{8}$&1&${\tiny \yng(1)}$&0&1& $R_S$ \\
$\lambda$ &$\mathbf{Adj.}$&1&1&0&0&$1$  \\ \hline
$\eta=\Lambda_{N_f,N_S}^b$&1&1&1&$2N_f$&$2N_S$&$6(R_f-1)+4(R_S-1) +10$  \\ \hline 
$M_{QQ}:=QQ$&1&$\tiny \yng(2)$&1&2&0&$2R_f$ \\
$M_{SS}:=SS$&1&1&$\tiny \yng(2)$&0&2&$2R_S$ \\
$P_{S3}:=SQ^3S$&1&1&${\tiny \yng(2)}$&3&2& $3R_f+2R_S$ \\
$P_{A1}:=SQS$&1&${\tiny \yng(1)}$&1&1&2&$R_f+2R_S$ \\
$P_{A2}:=SQ^2S$ &1&${\tiny \bar{ \yng(1)}}$&1&2&2& $2R_f+2R_S$ \\ \hline
$Z:=Y_1Y_2^2Y_3$&1&1&1&$-6$&$-4$& $-8  -6 (R_f-1) -4(R_S-1)=2-6R_f-4R_S$  \\ 
$Y:=\sqrt{Y_1 Z}$ $(\phi_1 \ge \phi_3)$ &1&1&1&$-3$&$-4$& $-5-3(R_f-1) -4(R_S-1)=2-3R_f-4R_S$  \\ \hline
  \end{tabular}}
  \end{center}\label{T32}
\end{table}

From the zero-mode counting of the Coulomb branch operators, we expected that there are two Coulomb branch directions un-lifted. One coordinate would be globally defined on the whole Weyl chamber, which was denoted by $Z$, and the other is defined on the region of $\phi_1 > \phi_3$, which is $Y$ or $Y_{spin}$.
From various consistency checks, we assume that these two directions are $Z$ and $Y$ in this case. Being different from the other examples, we cannot find any simple relation between them (we need to include at least a fractional power of the Higgs branch operators). Consequently, the two-dimensional coordinates are necessary for the quantum Coulomb branch of $(N_f,N_S)=(3,2)$. One can write down the superpotential consistent with all the symmetries listed in Table \ref{T32}.

\begin{align}
W  &= Z  \left( \det \, M_{QQ} \det \, M_{SS} -\det \, P_{S3}+  P_{A2}^2 M_{QQ} -\frac{1}{2} P_{A1}^2 M_{QQ}^2 \right)  \nonumber  \\
& \qquad +Y \left( P_{A1}P_{A2} -M_{SS} P_{S3} \right)+\eta Z,
\end{align}
where the last term appears when we put a theory on $\mathbb{S}^1 \times \mathbb{R}^3$ and it is absent in a 3d discussion.
We can check the validity of this s-confinement phase in various ways. First, remember that the 4d superpotential for $(N_f,N_S) =(3,2) $ takes the following form \cite{Cho:1997kr},
\begin{align}
W_{4d}^{(N_f,N_S)=(3,2)} &=X_1 \left( \det \, M_{QQ} \det \, M_{SS} - \det \, P_{S3}+  P_{A2}^2 M_{QQ} -\frac{1}{2} P_{A1}^2 M_{QQ}^2 +\eta \right) \nonumber \\
&\qquad +X_2 \left( P_{A1}P_{A2} - M_{SS} P_{S3} \right)
\end{align}
and this is easily reproduced by integrating out the two Coulomb branch operators. Second, we consider the Higgs branch along which the gauge group is broken to $G_2$. This can be achieved by higgsing the spinorial matter, let's say $\braket{M_{SS,22}}=v^2$. In order to properly obtain the $G_2$ superpotential we have to rename the fields as
\begin{gather}
M_{QQ} =: M^{G_2}_{ij} ~(i,j \le 3),~~~~M_{SS,11} =: M^{G_2}_{44},\\
P_{A,1}=:2v M^{G_2}_{i4},~~~P_{A,2}=: v B_{G_2}^i \\
P_{S3,12}=:v F_{G_2},~~~P_{S3,11}=: \sum_{i=1,2,3} 2B^i_{G_2} M_{i4}^{G2} + B^4_{G_2} M_{44}^{G_2},~~~P_{S3,22}=:-v^2B_{G2}^4,
\end{gather}
where $Y$ and $P_{S3,11}$ become massive and integrated out. By substituting these expressions into the superpotential, we arrive at the $G_2$ superpotential \cite{Nii:2017npz}.

Next, let us study another Higgs branch $\braket{M_{QQ,33}}=v^2$ along which the $Spin(7)$ group is broken to $SU(4)$. The low-energy theory becomes a 3d $\mathcal{N}=2$ $SU(4)$ gauge theory with two antisymmetric matters and with two (anti-)fundamental flavors. This was studied in \cite{Csaki:2014cwa} (see also \cite{Nii:2016jzi}). Table \ref{Tsu42anti} shows the matter contents, moduli coordinates and their quantum numbers. This theory has a two-dimensional Coulomb branch parametrized by $Y$ and $\tilde{Y}$. These two monopole operators correspond to the breaking $SU(4) \rightarrow SU(2) \times U(1) \times U(1)$ and $SU(4) \rightarrow SU(2) \times SU(2) \times U(1)$ respectively. This theory is known to be s-confining and the effective superpotential becomes
\begin{align}
W= Y(T^2 \det M_0 +TB \bar{B} +\det M_2 ) +\tilde{Y} (M_0M_2 +B \bar{B}) \label{Wsu4},
\end{align}
where we neglected the relative coefficients for simplicity.
 \begin{table}[H]\caption{$SU(4)$ with 2 ${\tiny \protect\yng(1,1)}$ and 2 $({\tiny \protect\yng(1)} +\protect\overline{{\tiny \protect\yng(1)}})$} 
\begin{center}
\scalebox{1}{
  \begin{tabular}{|c||c||c|c|c|c|c|c|c| } \hline
&$SU(4)$&$SU(2)$& $SU(2)$&$SU(2)$&$U(1)$&$U(1)$&$U(1)$&$U(1)_R$ \\ \hline
$A$&${\tiny \yng(1,1)}$&${\tiny \yng(1)}$&1&1&1&0&0 &0\\
$Q$&${\tiny \yng(1)}$&1&${\tiny \yng(1)}$&1&0&1&1&0 \\
$\tilde{Q}$&${\tiny \overline{\yng(1)}}$&1&1&${\tiny \yng(1)}$&0&1&$-1$&0 \\ \hline
$T:=A^2$&1&${\tiny \yng(2)}$&1&1&2&0&0&0 \\
$M_0:=Q \tilde{Q}$&1&1&${\tiny \yng(1)}$&${\tiny \yng(1)}$&0&2&0&0 \\
$M_2:=Q A^2\tilde{Q}$&1&1&${\tiny \yng(1)}$&${\tiny \yng(1)}$&2&2&0&0 \\
$B:=AQ^2$&1&${\tiny \yng(1)}$&1&1&1&2&2&0 \\
$\bar{B}:=A\tilde{Q}^2$  &1&${\tiny \yng(1)}$&1&1&1&2&$-2$&0 \\ \hline
$Y$&1&1&1&1&$-4$&$-4$&0&2 \\
$\tilde{Y}$&1&1&1&1&$-2$&$-4$&0&2 \\ \hline
  \end{tabular}}
  \end{center}\label{Tsu42anti}
\end{table}
\noindent Since the global symmetries are enhanced to $SU(2) \times SU(2) \times SU(2) $, the fields are decomposed into
\begin{gather}
M_{QQ,ij \le 2} =: T, ~~M_{SS,ii} =:M_{0 ii},~~ M_{SS,12} =: \frac{M_{0,12} +M_{0,21}}{2},  \\
P_{S3,ii} =: vM_{2,ii}, ~~P_{S3,12}=: \frac{v(M_{2,12} +M_{2,21}) }{2},\\
P_{A,1i}=: \frac{B_i+\bar{B}_i}{\sqrt{2}}~(i=1,2),~~P_{A,1,3} = \frac{v(M_{0,12} -M_{0,21})}{\sqrt{2}}\\
P_{A,2,i} =: \frac{v\epsilon^{ij}(B-\bar{B})_j }{ \sqrt{2}},~~P_{A,2,3} =:  \frac{M_{2,12} -M_{2,21}}{\sqrt{2}}.
\end{gather}
By substituting these expressions into the superpotential, we reproduce the superpotential \eqref{Wsu4}.

\subsubsection*{Superconformal Indices}
As another non-trivial check of our analysis, we study the superconformal indices of the 3d $\mathcal{N}=2$ $Spin(7)$ gauge theory with $(N_f,N_S)=(3,2)$. Since the dual description has no gauge group, the index is simple and expanded as

\tiny
\begin{align}
I_{dual}&=1+x^{1/3} \left(\frac{1}{t^6 u^4}+6 t^2+3 u^2\right)+3 t u^2 \sqrt{x}+x^{2/3} \left(\frac{1}{t^{12} u^8}+\frac{3}{t^6 u^2}+\frac{6}{t^4 u^4}+21 t^4+21 t^2 u^2+6 u^4\right) \nonumber \\
&\quad +x^{5/6} \left(\frac{3}{t^5 u^2}+\frac{1}{t^3 u^4}+21 t^3 u^2+9 t u^4\right)+x \left(\frac{1}{t^{18} u^{12}}+\frac{3}{t^{12} u^6}+\frac{6}{t^{10} u^8}+56 t^6+\frac{6}{t^6}+81 t^4 u^2+\frac{21}{t^4 u^2}+51 t^2 u^4+\frac{21}{t^2 u^4}+10 u^6\right) \nonumber \\
& \quad +x^{7/6} \left(\frac{3}{t^{11} u^6}+\frac{1}{t^9 u^8}+81 t^5 u^2+\frac{9}{t^5}+71 t^3 u^4+\frac{21}{t^3 u^2}+18 t u^6+\frac{6}{t u^4}\right) +x^{4/3} \biggl(\frac{1}{t^{24} u^{16}}+\frac{3}{t^{18} u^{10}}+\frac{6}{t^{16} u^{12}} \nonumber \\
&\qquad \quad +\frac{6}{t^{12} u^4}+\frac{21}{t^{10} u^6}+\frac{21}{t^8 u^8}+126 t^8+231 t^6 u^2+\frac{10 u^2}{t^6}+231 t^4 u^4+\frac{51}{t^4}+96 t^2 u^6+\frac{81}{t^2 u^2}+15 u^8+\frac{56}{u^4}\biggr)+\cdots, \label{dual}
\end{align}
\normalsize

\noindent where $t$ and $u$ are the fugacities for the $U(1)_Q$ and $U(1)_S$ symmetries. We set $R_f=R_S=\frac{1}{6}$ for simplicity.

Next, let us consider the index on the electric side for each GNO charge. We start with the sector with zero GNO charge.

\tiny
\begin{align}
I_{electric}^{(0,0,0)} &=1+x^{1/3} \left(6 t^2+3 u^2\right)+3 t u^2 \sqrt{x}+x^{2/3} \left(21 t^4+21 t^2 u^2+6 u^4\right)+x^{5/6} \left(21 t^3 u^2+9 t u^4\right)+x \left(56 t^6+81 t^4 u^2+51 t^2 u^4+10 u^6\right) \nonumber \\
&+x^{7/6} \left(81 t^5 u^2+71 t^3 u^4+18 t u^6\right)+x^{4/3} \left(126 t^8+231 t^6 u^2+231 t^4 u^4+96 t^2 u^6+15 u^8\right)+x^{3/2} \left(231 t^7 u^2+300 t^5 u^4+160 t^3 u^6+30 t u^8\right)\nonumber \\
&+x^{5/3} \left(252 t^{10}+546 t^8 u^2+746 t^6 u^4+486 t^4 u^6+156 t^2 u^8+21 u^{10}\right)+3 t x^{11/6} \left(182 t^8 u^2+305 t^6 u^4+250 t^4 u^6+96 t^2 u^8+15 u^{10}\right) +\cdots
\end{align}
\normalsize

\noindent This sector contains only the Higgs branch operators. $M_{QQ}, M_{SS}, P_{S3}, P_{A1}$ and $ P_{A2}$ appear as $6t^2 x^{1/3}, 3 u^2 x^{1/3}, 3t^3u^2 x^{5/6}, 3 t u^2 x^{1/2}$ and $3 t^2 u^2 x^{2/3} $ respectively. These are consistent with our analysis (see Table \ref{T32}). The next contribution is a sector with a GNO charges $\left(\frac{1}{2},\frac{1}{2},0 \right)$.

\tiny
\begin{align}
I_{electric}^{\left(\frac{1}{2},\frac{1}{2},0 \right)} &=\frac{x^{1/3}}{t^6 u^4}+x^{2/3} \left(\frac{3}{t^6 u^2}+\frac{6}{t^4 u^4}\right)+x^{5/6} \left(\frac{3}{t^5 u^2}+\frac{1}{t^3 u^4}\right)+x \left(\frac{6}{t^6}+\frac{21}{t^4 u^2}+\frac{21}{t^2 u^4}\right)+x^{7/6} \left(\frac{9}{t^5}+\frac{21}{t^3 u^2}+\frac{6}{t u^4}\right)\nonumber \\ 
&\quad+x^{4/3} \left(\frac{10 u^2}{t^6}+\frac{51}{t^4}+\frac{81}{t^2 u^2}+\frac{56}{u^4}\right)+x^{3/2} \left(\frac{18 u^2}{t^5}+\frac{68}{t^3}+\frac{21 t}{u^4}+\frac{81}{t u^2}\right)+x^{5/3} \left(\frac{15 u^4}{t^6}+\frac{96 u^2}{t^4}+\frac{126 t^2}{u^4}+\frac{216}{t^2}+\frac{231}{u^2}\right)\nonumber \\
&\quad +x^{11/6} \left(\frac{30 u^4}{t^5}+\frac{56 t^3}{u^4}+\frac{152 u^2}{t^3}+\frac{231 t}{u^2}+\frac{270}{t}\right)+x^2 \left(\frac{21 u^6}{t^6}+\frac{156 u^4}{t^4}+\frac{252 t^4}{u^4}+\frac{441 u^2}{t^2}+\frac{546 t^2}{u^2}+650\right)+\cdots
\end{align}
\normalsize

\noindent This sector contains two Coulomb branch operators. From Table \ref{T32}, the first term $\frac{x^{1/3}}{t^6 u^4}$ is identified with the operator $Z$. The other operator $Y$ also appears in this sector as $\frac{x^{5/6}}{t^3 u^4}$. The GNO charge (or the vev of $Z$) breaks the gauge group to $Spin(3) \times U(2)$. Under this breaking, the vector matters supply a $\mathbf{3}$ representation of $Spin(3)$. Consequently, $Y$ is understood also as $Y \sim Z t^3 x^{1/2} \sim Z Q^3$, where $ZQ^3$ is regarded as the product with the monopole and the $Spin(3)$ baryon. We cannot UV-complete $Q^3$ into a gauge invariant operator in terms of the UV elementary fields. Therefore, we need two monopole operators for the quantum Coulomb moduli. Up to $O(x^2)$, we have to also include the following sectors.

\tiny
\begin{align}
I_{electric}^{(1,1,0)} &=\frac{x^{2/3}}{t^{12} u^8}+x \left(\frac{3}{t^{12} u^6}+\frac{6}{t^{10} u^8}\right)+x^{7/6} \left(\frac{3}{t^{11} u^6}+\frac{1}{t^9 u^8}\right)+x^{4/3} \left(\frac{6}{t^{12} u^4}+\frac{21}{t^{10} u^6}+\frac{21}{t^8 u^8}\right)+x^{3/2} \left(\frac{9}{t^{11} u^4}+\frac{21}{t^9 u^6}+\frac{6}{t^7 u^8}\right) \nonumber \\
&+x^{5/3} \left(\frac{10}{t^{12} u^2}+\frac{51}{t^{10} u^4}+\frac{81}{t^8 u^6}+\frac{56}{t^6 u^8}\right)+x^{11/6} \left(\frac{18}{t^{11} u^2}+\frac{68}{t^9 u^4}+\frac{81}{t^7 u^6}+\frac{21}{t^5 u^8}\right)+x^2 \left(\frac{15}{t^{12}}+\frac{96}{t^{10} u^2}+\frac{216}{t^8 u^4}+\frac{231}{t^6 u^6}+\frac{126}{t^4 u^8}\right)+\cdots \\
I_{electric}^{\left(\frac{3}{2},\frac{3}{2},0 \right)} &=\frac{x}{t^{18} u^{12}}+x^{4/3} \left(\frac{3}{t^{18} u^{10}}+\frac{6}{t^{16} u^{12}}\right)+x^{3/2} \left(\frac{3}{t^{17} u^{10}}+\frac{1}{t^{15} u^{12}}\right)+x^{5/3} \left(\frac{6}{t^{18} u^8}+\frac{21}{t^{16} u^{10}}+\frac{21}{t^{14} u^{12}}\right) \nonumber \\
&+x^{11/6} \left(\frac{9}{t^{17} u^8}+\frac{21}{t^{15} u^{10}}+\frac{6}{t^{13} u^{12}}\right)+x^2 \left(\frac{10}{t^{18} u^6}+\frac{51}{t^{16} u^8}+\frac{81}{t^{14} u^{10}}+\frac{56}{t^{12} u^{12}}\right)+\cdots \\
I_{electric}^{(2,2,0)}&=\frac{x^{4/3}}{t^{24} u^{16}}+x^{5/3} \left(\frac{3}{t^{24} u^{14}}+\frac{6}{t^{22} u^{16}}\right)+x^{11/6} \left(\frac{3}{t^{23} u^{14}}+\frac{1}{t^{21} u^{16}}\right)+x^2 \left(\frac{6}{t^{24} u^{12}}+\frac{21}{t^{22} u^{14}}+\frac{21}{t^{20} u^{16}}\right)+\cdots \\
I_{electric}^{\left(\frac{5}{2},\frac{5}{2},0 \right)}&=\frac{x^{5/3}}{t^{30} u^{20}}+\frac{3 x^2 \left(2 t^2+u^2\right)}{t^{30} u^{20}} +\cdots,~~I_{electric}^{(3,3,0)}=\frac{x^2}{t^{36} u^{24}} +\cdots,~~
I_{electric}^{(1,0,0)} = \frac{x^{5/3}}{t^6 u^8}+\frac{6 x^2}{t^4 u^8} +\cdots,~~
I_{electric}^{(3/2,1/2,0)} =\frac{x^2}{t^{12} u^{12}}+\cdots. 
\end{align}
\normalsize

\noindent These indices are consistent with the index of the dual side \eqref{dual}.

\normalsize

\subsection{$(N_f,N_S )=(4,1) $}
In this subsection, we will investigate the 3d $\mathcal{N}=2$ $Spin(7)$ gauge theory with four vectors and one spinor.
In order to describe the Higgs branch of the moduli space, we need to define the following gauge invariant operators 
\begin{gather}
M_{QQ}:=QQ,~~~M_{SS}:=SS  \\
P:=S Q^3 S,~~~R:=S Q^4 S .
\end{gather}
Notice that only the symmetric product of the spinor is available. The theory has the $SU(4) \times U(1)_Q \times U(1)_S \times U(1)_R$ global symmetries. Table \ref{T41} shows the quantum numbers of the moduli coordinates.

 \begin{table}[H]\caption{Quantum numbers for $(N_f,N_S )=(4,1) $} 
\begin{center}
\scalebox{0.88}{
  \begin{tabular}{|c||c||c|c|c| } \hline
  &$SU(4)_Q$&$U(1)_Q$&$U(1)_S$&$U(1)_R$ \\ \hline
$\eta=\Lambda_{N_f,N_S}^b$&1&$8$&$2$&$8(R_f-1)+2(R_S-1) +10=8R_f+2R_S$  \\ \hline 
$M_{QQ}:=QQ$&$\tiny \yng(2)$&2&0&$2R_f$ \\
$M_{SS}:=SS$&1&0&2&$2R_S$ \\
$P:=SQ^3S$&${\tiny \bar{\yng(1)}}$&3&2& $3R_f+2R_S$ \\
$R:=SQ^4S$&1&4&2& $4R_f+2R_S$\\ \hline
$Z:=Y_1Y_2^2Y_3$&1&$-8$&$-2$& $-8 -2 (R_S-1) -8 (R_f-1)=2-8R_f-2R_S$  \\ 
$Y_{spin}:= Y_1 Z$ $(\phi_1 \ge \phi_3)$&1&$-8$&$-4$& $-10 -8(R_f-1)-4(R_S-1)=2-8R_f -4R_S$ \\ \hline
  \end{tabular}}
  \end{center}\label{T41}
\end{table}

From the analysis of the Coulomb branch corresponding to the semi-classical monopoles, one might expect that two-dimensional subspace of the classical Coulomb moduli remains flat and these are parametrized by $Z$ and $Y_{spin}$. In this case, however one can identify these two Coulomb branch operators as $Z \sim Y_{spin}M_{SS}$. Therefore it is plausible to expect that  the quantum Coulomb branch is one-dimensional. The superpotential consistent with all the symmetries takes
\begin{align}
W 
     &=Y_{spin} [M_{SS}^2 \det \, M_{QQ} +P^2 M_{QQ} - R^2 ]  +\eta Y_{spin } M_{SS}, \label{W41}
\end{align}
where the term proportional to $\eta$ is generated by a KK-monopole and absent in a 3d limit. Originally the KK-monopole contribution is $\eta Z$ but now it is expressed in terms of $Y_{spin}$. We can easily check the parity anomaly matching between the UV theory and the IR description \eqref{W41}. One might consider that the quantum Coulomb branch is described by $Y$ instead of $Y_{spin}$. However, in this case, we cannot satisfy the parity anomaly matching for $k_{U(1)_R U(1)_R}$.
By integrating out the Coulomb branch $Y_{spin}$, we reproduce the 4d result with a single quantum constraint \cite{Cho:1997kr}
\begin{align}
M_{SS}^2 \det \, M_{QQ} +P^2 M_{QQ} - R^2+\eta M_{SS}=0.
\end{align}
Therefore, the identification, $Z \sim Y_{spin}M_{SS}$, properly reduces the 3d result to the 4d constraint. Let us check the complex mass deformation for the spinorial matter, which leads to the 3d $\mathcal{N}=2$ $Spin(7)$ gauge theory with four vector matters. The superpotential becomes
\begin{align}
W=Y_{spin} [M_{SS}^2 \det \, M_{QQ} +P^2 M_{QQ} - R^2 ]  +m M_{SS}
\end{align}
and the equations of motion for $M_{SS},P$ and $R$ are
\begin{gather}
m+2Y_{spin} M_{SS} \det M_{QQ} =0, \\
Y_{spin}PM_{QQ} =0, \\
RY_{spin} =0,
\end{gather}
which lead to $P^i=R=0$ and $M_{SS}$ is integrated out. The low-energy superpotential results in
\begin{align}
W=\frac{1}{Y_{spin} \det M_{QQ}}.
\end{align}
This is consistent with the observation in \cite{Aharony:2011ci} with modification of the Coulomb branch operator. This difference is due to the fact that we deal with not an $SO(7)$ group but a $Spin(7)$ group.

Next, we will test the Higgs branch. When the spinor gets a vev $\braket{M_{SS}}=v^2$, the gauge group is broken to $G_2$. The low-energy limit becomes a 3d $\mathcal{N}=2$ $G_2$ gauge theory with four fundamentals from the vector matters.  Under the breaking we have the following identification between the $Spin(7)$ and $G_2$ theories
\begin{align}
P^i=:v^2 B_{G_2}^i,~~R=:vF_{G_2},~~Y_{spin }v^2 =:Z_{G_2}.
\end{align}
The superpotential reduces to
\begin{align}
W=Z_{G_2} \left[ \det M_{QQ} -F^2 +BM_{QQ}B \right],
\end{align}
which is precisely the $G_2$ superpotential observed in \cite{Nii:2017npz}.

Let us consider the different direction of the Higgs branch $\braket{M_{QQ,44}} =v^2$, along which the gauge group is broken as $Spin(7) \rightarrow SU(4)$. The low-energy theory becomes a 3d $\mathcal{N}=2$ $SU(4)$ gauge theory with three antisymmetric matters and one (anti-)fundamental flavor.
Since the UV theory is s-confining, the low-energy $SU(4)$ theory is also confining. We can directly show that this theory indeed exhibits an s-confinement phase. Table \ref{su43anti} shows the matter contents of the $SU(4)$ theory and their quantum numbers.

 \begin{table}[H]\caption{$SU(4)$ with 3 ${\tiny \protect\yng(1,1)}$ and $({\tiny \protect\yng(1)}+{\tiny \overline{\protect\yng(1)}})$} 
\begin{center}
\scalebox{1}{
  \begin{tabular}{|c||c||c|c|c|c|c| } \hline
  &$SU(4)$&$SU(3)$&$U(1)$&$U(1)_B$&$U(1)_A$&$U(1)_R$ \\ \hline
$A$  &${\tiny \yng(1,1)}$&${\tiny \yng(1)}$&1&0&0&$R_A$ \\
$Q$&${\tiny \yng(1)}$&1&0&1&1&$R_Q$ \\
$\tilde{Q}$&${\tiny \overline{\yng(1)}}$&1&0&1&$-1$&$R_Q$ \\ \hline
$T:=A^2=\mathrm{Pf}\,A$&1&${\tiny \yng(2)}$&2&0&0&$2R_A$ \\
$M_0:=Q\tilde{Q}$&1&1&0&2&0&$2R_Q$ \\
$M_2:=QA^2\tilde{Q}$&1&${\tiny \overline{\yng(1)}}$&2&2&0&$2R_A+2R_Q$ \\
$B:=QA^3Q$&1&1&3&2&2&$3R_A+2R_Q$  \\
$\bar{B}:=\tilde{Q}A^3\tilde{Q}$&1&1&3&2&$-2$&$3R_A+2R_Q$ \\ \hline
$Y$&1&1&$-6$&$-2$&0&$2-2R_Q-6R_A$ \\
$\hat{Y}$&1&1&$-6$&$-4$&$0$&$2-4R_Q-6R_A$ \\ \hline
  \end{tabular}}
  \end{center}\label{su43anti}
\end{table}
From the classical analysis of the $SU(4)$ Coulomb brach (see \cite{Csaki:2014cwa,Amariti:2015kha}), one might expect that there are two types of Coulomb branch corresponding to
\begin{align}
Y \leftrightarrow \begin{pmatrix}
\sigma & && \\
 &0&& \\
 &&0& \\
 &&& -\sigma\\
\end{pmatrix}
 ,~~~~~~ \hat{Y}  \leftrightarrow \begin{pmatrix}
\sigma & && \\
 &\sigma&& \\
 &&-\sigma& \\
 &&& -\sigma\\
\end{pmatrix}.
\end{align}
However, Table \ref{su43anti} suggests that these two variables are related as $Y \sim \hat{Y}M_0.$ Consequently the quantum Coulomb branch becomes one-dimensional. We obtain the confining superpotential
\begin{align}
W= \hat{Y}(T^3M_0^2+TM_2^2+B\bar{B}). \label{Wsu431}
\end{align}
One can flow to this superpotential also from the UV description of \eqref{W41}. In order to show this, we have to rename the fields as follows  
\begin{gather}
M_{QQ,ij} =: T ~~(i,j=1,2,3),~~~M_{SS}=:M_0\\
P^{i=1,2,3} =:v M_2,~~~P^{i=4}=:\frac{B+\bar{B}}{2},~~~R=: \frac{v(B-\bar{B})}{2}.
\end{gather}
By substituting these expressions, we reproduce the superpotential \eqref{Wsu431}.

\subsubsection*{Superconformal Indices of $SU(4)$ with 3 ${\tiny \protect\yng(1,1)}$ and $({\tiny \protect\yng(1)}+{\tiny \overline{\protect\yng(1)}})$}

Let us first study the superconformal indices of the 3d $\mathcal{N}=2$ $SU(4)$ gauge theory with three antisymmetric matters and with one (anti-)fundamental flavor. Since the theory is s-confining, the index must be equivalent to the index of the dual description with $T,M_0,M_2,B,\bar{B}$ and $\hat{Y}$ (not including $Y$). The full index (or the index on the magnetic side) is

\scriptsize
\begin{align}
I_{mag}&=1+x^{1/3} \left(\frac{1}{t^4 u^6}+t^2+6 u^2\right)+x^{2/3} \left(\frac{1}{t^8 u^{12}}+\frac{6}{t^4 u^4}+t^4+\frac{1}{t^2 u^6}+9 t^2 u^2+21 u^4\right)+2 t^2 u^3 x^{5/6} \nonumber \\
&\quad +x \left(\frac{1}{t^{12} u^{18}}+\frac{6}{t^8 u^{10}}+\frac{1}{t^6 u^{12}}+t^6+9 t^4 u^2+\frac{21}{t^4 u^2}+39 t^2 u^4+\frac{9}{t^2 u^4}+56 u^6+\frac{1}{u^6}\right)+2 t^2 u^3 x^{7/6} \left(t^2+6 u^2\right)\nonumber \\
&\quad +x^{4/3} \biggl(\frac{1}{t^{16} u^{24}}+\frac{6}{t^{12} u^{16}}+\frac{1}{t^{10} u^{18}}+\frac{21}{t^8 u^8}+t^8+\frac{9}{t^6 u^{10}} \nonumber \\
&\qquad \qquad +9 t^6 u^2+\frac{\frac{1}{u^{12}}+56}{t^4}+45 t^4 u^4+t^2 \left(119 u^6+\frac{1}{u^6}\right)+\frac{36}{t^2 u^2}+126 u^8+\frac{9}{u^4}\biggr)+\cdots, \label{ISUF}
\end{align}
\normalsize

\noindent where $t$ represents the $U(1)$ charge of the (anti-)fundamental matters and $u$ counts the anti-symmetric matters. We set $R_A=R_Q=\frac{1}{6}$ for simplicity.

Next, we show the index for each GNO charge. The important sectors are only listed below.

\scriptsize
\begin{align}
I_{electric}^{(0,0,0)} &=1+x^{1/3} \left(t^2+6 u^2\right)+x^{2/3} \left(t^4+9 t^2 u^2+21 u^4\right)+2 t^2 u^3 x^{5/6}+x \left(t^6+9 t^4 u^2+39 t^2 u^4+56 u^6\right) \nonumber \\
&+x^{7/6} \left(2 t^4 u^3+12 t^2 u^5\right)+x^{4/3} \left(t^8+9 t^6 u^2+45 t^4 u^4+119 t^2 u^6+126 u^8\right)+x^{3/2} \left(2 t^6 u^3+18 t^4 u^5+42 t^2 u^7\right) \nonumber \\
&+x^{5/3} \left(t^{10}+9 t^8 u^2+45 t^6 u^4+157 t^4 u^6+294 t^2 u^8+252 u^{10}\right)+x^{11/6} \left(2 t^8 u^3+18 t^6 u^5+78 t^4 u^7+112 t^2 u^9\right) \nonumber \\
& \qquad +x^2 \left(t^{12}+9 t^{10} u^2+45 t^8 u^4+167 t^6 u^6+432 t^4 u^8+630 t^2 u^{10}+462 u^{12}-11\right)+\cdots, \\
I_{electric}^{\left(\frac{1}{2},0,0 \right)} &=\frac{x^{2/3}}{t^2 u^6}+x \left(\frac{9}{t^2 u^4}+\frac{1}{u^6}\right)+x^{4/3} \left(\frac{t^2}{u^6}+\frac{36}{t^2 u^2}+\frac{9}{u^4}\right)+x^{5/3} \left(\frac{t^4}{u^6}+\frac{9 t^2}{u^4}+\frac{100}{t^2}+\frac{36}{u^2}\right) \nonumber \\
&+x^2 \left(\frac{t^6}{u^6}+\frac{9 t^4}{u^4}+\frac{36 t^2}{u^2}+\frac{225 u^2}{t^2}+100\right)+\cdots, \\
I_{electric}^{\left(\frac{1}{2},\frac{1}{2},-\frac{1}{2} \right)} &=\frac{x^{1/3}}{t^4 u^6}+\frac{6 x^{2/3}}{t^4 u^4}+\frac{21 x}{t^4 u^2}+\frac{56 x^{4/3}}{t^4}+\frac{126 u^2 x^{5/3}}{t^4}+\frac{252 u^4 x^2}{t^4}+\cdots.
\end{align}
\normalsize

\noindent The sector with zero GNO charge contains the Higgs branch operators. The second term $x^{1/3} \left(t^2+6 u^2\right)$ represents the mesons $M_0$ and $T$. The third term $x^{2/3} \left(t^4+9 t^2 u^2+21 u^4\right)$ contains $M_0^2,T^2, M_0T$ and $M_2$. The fourth term corresponds to the baryonic operators $B$ and $\bar{B}$. 
The sector with a GNO charge $\left(\frac{1}{2},0,0 \right)$ classically represents the Coulomb branch operator $Y$ as $\frac{x^{2/3}}{t^2 u^6}$, which is not a quantum Coulomb branch operator. In this sector, the gauge group is broken to $SU(2) \times U(1) \times U(1)$. Therefore, the BPS scalar states are $Y M_0$ and $Y A^2$ where $M_0$ and $A^2$ are constructed from the fields not interacting with the monopole background. Hence $Y A^2$ contains the nine contributions $\frac{9x}{t^2 u^4}$ while $T:=A^2$ has six components. Quantum mechanically, these nine contributions are decomposed into $\hat{Y} M_0 T$ and $\hat{Y}M_2$. This observation is very consistent with our prediction $Y \sim \hat{Y} M_0$. 
The sector with a GNO charge $\left(\frac{1}{2},\frac{1}{2},-\frac{1}{2}  \right)$ contains the genuine Coulomb branch operator $\hat{Y}$ as $\frac{x^{1/3}}{t^4 u^6}$ and this is consistent with Table \ref{su43anti}. Since the gauge group is broken to $SU(2) \times SU(2) \times U(1)$ in this sector, we cannot take a product between $\hat{Y}$ and the (anti-)fundamental fields which are all charged under the $U(1)$. Therefore the proceeding terms are identified with $\hat{Y}T^n$. By summing up all the other sectors contributing to the lower orders in the index we reproduce the full index \eqref{ISUF}.

\subsubsection*{Superconformal Indices of $Spin(7)$ with $(N_f,N_c)=(4,1)$}
We also discuss the superconformal indices for the 3d $\mathcal{N}=2$ $Spin(7)$ theory with $(N_f,N_c)=(4,1)$. Since the theory is s-confining, the full index should be equivalent to the index of the dual description \eqref{W41} without the last term. The R-charges of the elementary chiral superfields are all set to be $R_f=R_S=\frac{1}{8}$. The full index is given by
 
\tiny
\begin{align}
I_{magnetic}^{(N_f,N_S)=(4,1)} &=1+x^{1/4} \left(10 t^2+u^2\right)+\sqrt{x} \left(\frac{1}{t^8 u^4}+55 t^4+10 t^2 u^2+u^4\right)+4 t^3 u^2 x^{5/8} \nonumber \\\
&\qquad +x^{3/4} \left(220 t^6+56 t^4 u^2+10 t^2 u^4+\frac{10 t^2+u^2}{t^8 u^4}+u^6\right)+4 t^3 u^2 x^{7/8} \left(10 t^2+u^2\right) \nonumber \\
&\qquad+x \left(\frac{1}{t^{16} u^8}+715 t^8+\frac{1}{t^8}+230 t^6 u^2+\frac{10}{t^6 u^2}+56 t^4 u^4+\frac{55}{t^4 u^4}+10 t^2 u^6+u^8\right) \nonumber \\
&\qquad +4 t^3 u^2 x^{9/8} \left(\frac{1}{t^8 u^4}+55 t^4+10 t^2 u^2+u^4\right) \nonumber \\
&\qquad+x^{5/4} \left(\frac{1}{t^{16} u^6}+\frac{10}{t^{14} u^8}+2002 t^{10}+770 t^8 u^2+\frac{u^2}{t^8}+240 t^6 u^4+\frac{10}{t^6}+56 t^4 u^6+\frac{55}{t^4 u^2}+10 t^2 u^8+\frac{220}{t^2 u^4}+u^{10}\right) +\cdots, \label{I41m}
\end{align}
\normalsize

\noindent where $t$ and $u$ are the fugacities for the $U(1)_Q$ and $U(1)_S$ symmetries.

The index is decomposed into the sectors with different GNO charges on the electric side.
Since the $Spin (7)$ gauge group is considered, the following sectors are necessary up to $O(x^{5/4})$.

\scriptsize
\begin{align}
I_{electric}^{(0,0,0)} &= 1+x^{1/4} \left(10 t^2+u^2\right)+\sqrt{x} \left(55 t^4+10 t^2 u^2+u^4\right)+4 t^3 u^2 x^{5/8}+x^{3/4} \left(220 t^6+56 t^4 u^2+10 t^2 u^4+u^6\right) \nonumber \\
&\qquad +x^{7/8} \left(40 t^5 u^2+4 t^3 u^4\right)+x \left(715 t^8+230 t^6 u^2+56 t^4 u^4+10 t^2 u^6+u^8\right) +4 x^{9/8} \left(55 t^7 u^2+10 t^5 u^4+t^3 u^6\right) \nonumber \\
&\qquad+ x^{5/4} \left(2002 t^{10}+770 t^8 u^2+240 t^6 u^4+56 t^4 u^6+10 t^2 u^8+u^{10}\right) +\cdots \\
I_{electric}^{ \left( \frac{1}{2},\frac{1}{2},0   \right)} &=\frac{x^{3/4}}{t^8 u^2}+x \left(\frac{1}{t^8}+\frac{10}{t^6 u^2}\right)+\frac{4 x^{9/8}}{t^5 u^2}+x^{5/4} \left(\frac{u^2}{t^8} +\frac{10}{t^6}+\frac{55}{t^4 u^2}\right) +\cdots, \\
I_{electric}^{(1,0,0)} &=\frac{\sqrt{x}}{t^8 u^4}+\frac{10 x^{3/4}}{t^6 u^4}+\frac{55 x}{t^4 u^4}+\frac{220 x^{5/4}}{t^2 u^4} +\cdots, \\
I_{electric}^{(2,0,0)} &=\frac{x}{t^{16} u^8}+\frac{10 x^{5/4}}{t^{14} u^8}+\cdots,~~~~~~~~~~~~~~~~~~~~~~~~~~~~~~~~~~~
I_{electric}^{\left( \frac{3}{2} ,\frac{1}{2},0   \right)} =\frac{x^{5/4}}{t^{16} u^6} +\cdots.
\end{align}
\normalsize

\noindent The summation of these indices precisely matches the index \eqref{I41m} on the magnetic side. The index with a GNO charge $(1,0,0)$ explains the monopole operator $Y_{spin}$ whose R-chrage is $\frac{1}{2}$. This sector breaks the gauge group to $Spin(5) \times U(1)$. The spinor matters are all charged under this $U(1)$. Therefore, the BPS scalar states do not contain the spinorial fields on the $Y_{spin}$ background. The second term $\frac{10 x^{3/4}}{t^6 u^4}$ only includes $Y_{spin} M_{QQ}$. However one can generally consider the products of the chiral superfields $Y_{spin}$ and $M_{SS}$ in the chiral ring. These are contained in the sector with a GNO chaerge $\left( \frac{1}{2},\frac{1}{2},0   \right)$. This sector corresponds to the operator $Z \sim Y_{spin} M_{SS}$. The second term $x \left(\frac{1}{t^8}+\frac{10}{t^6 u^2}\right)$ is regarded as $Z(M_{SS}+M_{QQ}) \sim Y_{spin}(M_{SS}^2 +M_{SS}M_{QQ}) $. The third term $\frac{4 x^{9/8}}{t^5 u^2}$ corresponds to $ZQ^3 \sim Y_{spin}SSQ^3 \sim Y_{spin} P$. This is again consistent with our prediction $Z \sim Y_{spin} M_{SS}$.

\section{Summary and Discussion}
In this paper, we investigated the 3d $\mathcal{N}=2$ supersymmetric $Spin(7)$ gauge theories with spinorial and vectorial matters. The 3d $\mathcal{N}=2$ $Spin(7)$ gauge theory with only the spinor matters has the one-dimensional (quantum) Coulomb branch parametrized by $Z$. For $N_S \le 3$, we found no stable SUSY vacuum. For $N_S=4$, the Higgs branch and the Coulomb branch are merged. For $N_S=5$, the theory is s-confining.  For the theory with both spinors and vectors, the Coulomb branch becomes two-dimensional at least semi-classically and needs two coordinates $Z$ and $Y$ (or $Y_{spin}$). However, sometimes we can relate these two coordinates quantum-mechanically by taking the product of the Higgs and Coulomb branch coordinates. If this is possible, the Coulomb branch becomes one-dimensional. Especially we focused on the s-confinement phases which appear for $(N_f,N_S)=(0,5),(1,4),(2,3),(3,2)$ and $(4,1)$. We found and tested various s-confinement phases for the $Spin(7)$ theories. As a byproduct, we could obtain the s-confinement phases for the 3d $\mathcal{N}=2$ $SU(4)$ gauge theories with $n$ anti-symmetric matters and with $4-n$ (anti-)fundamental flavors. For $n=1,3$, the s-confinement phases were not known in the literature. We also tested the validity of our analysis by computing the superconformal indices. The indices are perfectly consistent with our prediction on the Coulomb branch coordinates and also consistent with the s-confinement phases which we found.

In this paper, we expected that two-dimensional coordinates are semi-classically described by $Z$ and $Y$ (or $Z$ and $Y_{spin}$). Since the $Z$ coordinate is globally defined without depending on the sign of $\phi_1-\phi_3$, it is plausible to expect that $Z$ is necessary in any cases. However we could not find a priori way for choosing $Y$ or $Y_{spin}$ for the description of the remaining Coulomb branch. Just from various consistencies (including the SCI calculation, parity anomaly matching, deformations), we decided which one is more appropriate. For instance, $Z$ and $Y$ are presumably the natural coordinates for $(N_f,N_S )=(3,2)$ while $Z$ and $Y_{spin}$ are chosen for $(N_f,N_S )=(4,1)$ and $Z$ was equivalent to $Y_{spin} M_{SS}$. However, these decisions and reasoning were not conclusive. It would be nice if we gain a clear understanding of the quantum Coulomb branch.

It is interesting to study 3d $\mathcal{N}=2$ $Spin(N)$ $(N>7)$ theories with vector matters and with spinor matters. In the case of $Spin(2N)$ groups, two types of spinor representations are available. Hence the phase diagrams would be more richer than the $Spin(2N+1)$ cases. We will soon come back to this generalization elsewhere.

It is worth searching for Seiberg dual descriptions for the 3d $\mathcal{N}=2$ $Spin(7)$ gauge theories with spinorial matters. In 4d, the dual theory has an $SU(N_S-4)$ gauge group with $N_S$ anti-fundamental matters and with a matter in a symmetric representation. When we naively put a dual theory on a circle, the resulting Coulomb branch would be more than one-dimensional because a symmetric tensor divides the (classical) Coulomb branch. Furthermore the Coulomb branch operators are dressed by Higgs branch operators \cite{Amariti:2015kha} because the matter contents are ``chiral'' in a 4d sense. Deriving the 3d duality from the 4d duality becomes very complicated in this case. We don't have any simple 3d dual to the $Spin(7)$ now but would like to report some progresses along this direction in the near future.

\section*{Acknowledgments}
This work is supported by the Swiss National Science Foundation (SNF) under grant number PP00P2\_157571/1.

\bibliographystyle{ieeetr}
\bibliography{spin7theory}

\end{document}